\documentclass[dvips,aoas,preprint]{imsart}

\RequirePackage[OT1]{fontenc}
\RequirePackage{amsthm,amsmath}
\RequirePackage[numbers]{natbib}
\newtheorem{remark}{Remark}
\RequirePackage{multirow}
\RequirePackage{ctable}
\RequirePackage{apacite}

\startlocaldefs
\numberwithin{equation}{section}
\theoremstyle{plain}

\endlocaldefs

\begin{document}

\begin{frontmatter}

\title{Variable Selection for Survival Data with A Class of Adaptive Elastic Net Techniques}
\runtitle{Variable Selection for High-dimension Survival Data}


\author{\fnms{Md Hasinur Rahaman} \snm{Khan}\corref{t1}\ead[label=e1]{hasinur@isrt.ac.bd}}
\address{Applied Statistics\\Institute of Statistical Research and Training\\University of Dhaka\\Dhaka 1000\\\printead{e1}}
\affiliation{University of Dhaka}
\and
\author{\fnms{J. Ewart H.} \snm{Shaw}\ead[label=e2]{Ewart.Shaw@warwick.ac.uk}}
\address{Department of Statistics\\University of Warwick\\Coventry CV4 7AL\\\printead{e2}}
\affiliation{University of Warwick}


\runauthor{Khan, MHR and Shaw, JEH}

\begin{abstract}
The accelerated failure time (AFT) models have proved useful in many contexts, though heavy censoring (as for example in cancer survival) and high dimensionality (as for example in microarray data) cause difficulties for model fitting and model selection. We propose new approaches to variable selection for censored data, based on AFT models optimized using regularized weighted least squares. The regularized technique uses a mixture of $\ell_1$ and $\ell_2$ norm penalties under two proposed elastic net type approaches. One is the the adaptive elastic net and the other is weighted elastic net. The approaches extend the original approaches proposed by Ghosh [Technical Report (2007\nocite{ghosh:07:adapEnet}) PR no.~07-01], and Hong and Zhang [Math.~Model.~of Natu.~Phen.~\textbf{5} (2010) 115-133\nocite{Hon:Zhang:2010:weighted})] respectively. We also extend the two proposed approaches by adding censoring observations as constraints into their model optimization frameworks. The approaches are evaluated on microarray and by simulation. We compare the performance of these approaches with six other variable selection techniques--three are generally used for censored data and the other three are correlation-based greedy methods used for high-dimensional data.
\end{abstract}


\begin{keyword}
\kwd{Adaptive elastic net}
\kwd{AFT}
\kwd{Variable selection}
\kwd{Stute's weighted least squares}
\kwd{Weighted elastic net}
\end{keyword}

\end{frontmatter}

\section{Introduction}
The practical importance of variable selection is huge and well recognized in many disciplines, and has been the focus of much research.
%
Several variable selection techniques have been developed for linear regression models; some of these have been extended to censored survival data. The methods include stepwise selection [Peduzzi, Hardy and Holford~(1980\nocite{pedu:hard:hold:80:astep})], and penalized likelihood based techniques, such as Akaike's information criterion (AIC) [Akaike et al.~(1973\nocite{akai:73:infor})], bridge regression [Frank and Friedman (1993\nocite{fran:frie:93:astat})], least absolute shrinkage and selection operator (lasso) [Tibshirani (1996\nocite{tibs:96:regres})], Smoothly Clipped Absolute Deviation (SCAD) [Fan and Li (2001)\nocite{fan:li:01:varia}], least angle regression selection (LARS) [Efron et al.~(2004\nocite{efro:hast:john:tibs:04:least})], the elastic net [Zou and Hastie (2005\nocite{zou:hast:05:regul})], MM algorithms [Hunter and Li (2005\nocite{hunt:li:05:vaia})] that are based on extensions of the well-known class of EM algorithms, group lasso [Yuan and Lin (2006\nocite{yua:lin:06:model})], the Dantzig selector [Candes and Tao (2007\nocite{cand:tao:07:thedantzig})] that is based on a selector that minimizes the $\ell_1$ norm of the coefficients subject to a constraint on the error terms, and MC+ [Zhang (2010\nocite{zhan:10:near})] that is based on a minimax concave penalty and penalized linear unbiased selection. Stability selection as proposed in Meinshausen and B\"{u}hlmann (2010\nocite{Mein:Buh:10:stability}) is a variable selection technique that is based on subsampling in combination with (high-dimensional) selection algorithms. It is also used as a technique to improve variable selection performance for a range of selection methods.

Recently there has been a surge of interest in variable selection with ultra-high dimensional data. By ultra-high dimension, Fan and Lv (2008\nocite{Fan:Lv:08:Sure}) meant that the dimensionality grows exponentially in the sample size, i.e., $\log(p) = O(n^a)$ for some $a\in(0,\,1/2)$. The issue of high correlations among the variables for variable selection with ultra-high dimensional data has been dealt with using various greedy approaches. For linear regression with ultra-high dimensional datasets, Fan and Lv (2008\nocite{Fan:Lv:08:Sure}) proposed sure independence screening (SIS) based on marginal correlation ranking. B\"{u}hlmann, Kalisch and Maathuiset~(2010\nocite{Buh:kal:maat:10:varia}) proposed the PC-simple algorithm, that uses partial correlation to infer the association between each variable and the response conditional on other variables. Radchenko and James (2011\nocite{Radc:Jam:11:Improv}) proposed the forward-lasso adaptive shrinkage (FLASH) that includes the lasso and forward selection as special cases at two extreme ends. Cho and Fryzlewicz (2012\nocite{Cho:Fryz:12:HighDim}) proposed a \emph{tilting} procedure that provides an adaptive choice between the use of marginal correlation and tilted correlation for each variable, where the choice is made depending on the values of the hard-thresholded sample correlation of the design matrix.

The Cox model [Cox (1978\nocite{cox:72:reg})] with high--dimensional data has been the focus of many variable selection studies. For example, Tibshirani (1997\nocite{tibs:97:thelasso}) developed a regularized Cox regression by minimizing an $\ell_1$ lasso penalty to the partial likelihood, Faraggi and Simon (1998\nocite{far:sim:98:bay}) proposed a Bayesian variable selection method, Fan and Li (2002\nocite{fan:li:02:variable}) developed a non--concave penalized likelihood approach, Li and Luan (2003\nocite{li:luan:03:kernel}) used kernel transformations, and Gui and Li (2005\nocite{gui:li:05:pena}) introduced a threshold gradient descent regularization estimation method, Antoniadis, Fryzlewicz and Letue~(2010\nocite{Anton:Fry:Let:10:TheDantzig}) developed a variable selection approach for the Cox model based on the Dantzig selector.

There are also some variable selection studies for AFT models. For example, Huang et al.~(2006\nocite{hua:ma:xie:06:regu}) used the lasso regularization for estimation and variable selection in the AFT model based on the inverse probability of censoring. The lasso regularized Buckley--James method for the AFT model is investigated by Huang and Harrington (2005\nocite{Huang:Harri:05:itera}) and Datta, Le-Rademacher and Datta~(2007\nocite{dat:le:dat:07:pred}). Sha, Tadesse and Vannucci~(2006\nocite{sha:tad:van:06:bay}) developed a Bayesian variable selection approach. Variable selection using the elastic net is investigated in Wang et al.~(2008\nocite{wang:nan:zhu:beer:08:doubly}). Engler and Li (2009\nocite{eng:li:09:surv}), and Cai, Huang and Tian~(2009\nocite{cai:huan:tian:09:regula}) proposed variable selection using the lasso regularized rank based estimator. Huang and Ma (2010\nocite{huan:ma:10:variab}) used a bridge method for variable selection. Hu and Rao (2010\nocite{hu:rao:2010:sparse}) proposed a sparse penalization technique with censoring constraints. Recently, Khan and Shaw (2013\nocite{Kha:sha:13:varia}) proposed a variable selection technique for AFT model that is based on the synthesize of the Buckley--James method and the Dantzig selector.

In this paper, we consider variable selection methods for the AFT modeling of censored data, and propose new regularized Stute's Weighted Least Squares (SWLS) approaches. We introduce classes of elastic net type regularized variable selection techniques based on SWLS. The classes include an adaptive elastic net, a weighted elastic net and two extended versions that are carried out by introducing censoring constraints into the optimization function.

The rest of the paper is structured as follows. Section \ref{sec2} provides the regularized framework of SWLS. Section \ref{sec:swlseReguNEW} provides proposed variable selection methods including variable selection criteria and prediction formula. All the methods are demonstrated with two simulated examples in Section \ref{sec3} and with one microarray real data example in Section \ref{sec4}. In Section \ref{sec3} and \ref{sec4} we also compare the performance of the proposed approaches with three other variable selection approaches: the typical elastic net implemented for weighted data, the adaptive elastic net for censored data [Engler and Li (2009\nocite{eng:li:09:surv})], and a Bayesian approach [Sha, Tadesse and Vannucci~(2006\nocite{sha:tad:van:06:bay})] and three other correlation-based greedy variable selection methods generally used for high-dimensional data: sure independence screening [Fan and Lv (2008\nocite{Fan:Lv:08:Sure})], tilted correlation screening [Cho and Fryzlewicz (2012\nocite{Cho:Fryz:12:HighDim})], and PC-simple [B\"{u}hlmann, Kalisch and Maathuiset~(2010\nocite{Buh:kal:maat:10:varia})].

\section{Methodology}\label{sec2}

\subsection{Regularized SWLS (Stute's Weighted Least Squares) and Censoring Constraints}\label{sec:swlseRegu}
The objective function of Stute's weighted least squares for a typical AFT model is given by
\begin{equation}
\begin{array}{c}\arg\min\ \\ (\alpha,\beta) \end{array}\left[\frac{1}{2} \sum_{i=1}^{n}w_{i}\,(Y_{(i)}-\alpha- X_{(i)}^T\,\beta)^2\right],
\label{eq:wlse}
\end{equation}
where $Y_i$ is the log survival time for the $i$-th observation, $\mathbf{X}$ is the covariate vector, $\alpha$ is the intercept term, $\beta$ is the unknown $p\times 1$ vector of true regression coefficients, and $w_{i}$ are the Kaplan--Meier (K--M) weights that are obtained by
\begin{equation}
w_{1}=\frac{\delta_{(1)}}{n},~ w_{i}=\frac{\delta_{(i)}}{n-i+1}\prod_{j=1}^{i-1}\Big(\frac{n-j}{n-j+1}\Big)^{\delta_{(j)}},~~i=2,\cdots,n
\label{eq:kmweights}
\end{equation}

As discussed in Huang et al.~(2006\nocite{hua:ma:xie:06:regu}), the SWLS method is computationally more amenable to high--dimensional covariates than the B--J estimator [Buckley and James (1979\nocite{buc:jam:79:lin})] or a rank based estimator [e.g.,~Ying (1993\nocite{yin:93:alarge})]. This is because the OLS structure makes it computationally efficient to apply a regularized method in the AFT model. The method has rigorous theoretical justifications under reasonable assumptions.

In matrix notation the objective function of SWLS (\ref{eq:wlse}) is given by
\begin{equation}
\frac{1}{2}\,(Y-\alpha-X\beta)^T w \,(Y-\alpha-X\beta),
\label{eq:swlse1}
\end{equation}
where $w$ is the $n \times n$ diagonal weight matrix. Let the uncensored and censored data be subscripted by $u$ and $\bar{u}$ respectively. Thus the number of uncensored and censored observations are denoted by $n_{u}$ and $n_{\bar{u}}$, the predictor and response observations for censored data by $X_{\bar{u}}$ and $Y_{\bar{u}}$, and the unobserved true failure time for censored observation by $T_{\bar{u}}$. Since under right censoring $Y_{\bar{u}}<\log(T_{\bar{u}})$ that is equivalent to $Y_{\bar{u}}\leq \alpha +X_{\bar{u}}\,\beta$, can be added to the SWLS objective function (\ref{eq:swlse1}). These constraints are called censoring constraints [Hu and Rao (2010\nocite{hu:rao:2010:sparse})] which may be too stringent due to the random noise. This might suggest modifying the constraints to $Y_{\bar{u}}\leq \alpha +X_{\bar{u}}\,\beta+\xi$, where $\xi$ is a vector of non-negative values that measure the severities of violations of the constraints. The SWLS objective function now can be defined by
\begin{align}
L(\alpha,\,\beta)={}&\frac{1}{2}\,(Y_{u}-\alpha-X_{u}\,\beta)^T\,w_u\,(Y_{u}-\alpha-X_{u}\,\beta)+\frac{\lambda_0}{2n}\,\xi^T\xi,\nonumber\\
&\text{subject to}~Y_{\bar{u}}\leq \alpha +X_{\bar{u}}\,\beta+\xi,
\label{eq:swlse2}
\end{align}
where $\lambda_0$ is a positive value that accounts for the penalties of violations of constraints, and $n$ is included for scaling to match the $w_u$.

An intercept term $\alpha$ typically is included in the AFT model. However, for notational convenience, we can remove $\alpha$ by (weighted) standardisation of the predictors and response. The weighted means are defined by
\begin{equation*}
\bar{X}_{w}=\frac{\sum_{i=1}^n w_{i}X_{(i)}}{\sum_{i=1}^n
w_{i}},\qquad \bar{Y}_{w}=\frac{\sum_{i=1}^n
w_{i}Y_{(i)}}{\sum_{i=1}^n w_{i}}.
\end{equation*}
Then the adjusted predictors and responses are defined by
\begin{equation*}
X_{(i)}^w=(w_{i})^{1/2}(X_{(i)}-\bar{X}_{w}),\qquad Y_{(i)}^w=(w_{i})^{1/2}(Y_{(i)}-\bar{Y}_{w}).
\end{equation*}
For simplicity, we still use $X_{(i)}$ and $Y_{(i)}$ to denote the weighted and centered values and $(Y_{(i)}, \,\delta_{(i)}, \,X_{(i)})$ to denote the weighted data.

The objective function of SWLS (\ref{eq:swlse1}) therefore becomes
\begin{equation}
L(\beta)=\frac{1}{2}\,(Y_{u}-X_{u}\,\beta)^T (Y_{u}-X_{u}\,\beta).
\label{eq:swlse}
\end{equation}
So, it is easy to show that the SWLS in Equation (\ref{eq:wlse}) is equivalent to the OLS estimator without intercept on the weighted data with K$-$M weights. Unfortunately, OLS estimation does not perform well with variable selection, and is simply infeasible when $p>n$. Hence the need to introduce various regularized methods that improve OLS, such as
lasso [Tibshirani (1996\nocite{tibs:96:regres})], the elastic net [Zou and Hastie (2005\nocite{zou:hast:05:regul})], and the Dantzig selector [Candes and Tao (2007\nocite{cand:tao:07:thedantzig})]. Many of these regularized methods are developed for data where $p>n$ and the coefficients vector is sparse. The general frame of regularized WLS objective function is therefore defined by

\begin{equation}
L(\beta,\, \lambda)=\frac{1}{2}\,(Y_{u}-X_{u}\,\beta)^T (Y_{u}-X_{u}\,\beta) + \lambda\, \mbox{pen}(\beta),
\label{eq:regwlse}
\end{equation}
where $\lambda$ is the (scalar or vector) penalty parameter and the penalty quantity $\mbox{pen}(\beta)$ is set typically in a way so that it controls the complexity of the model. For example, the penalty $\mbox{pen}(\beta)$ for ridge, lasso and elastic net are defined as
\begin{equation*}
\sum_{j=1}^p\beta_j^2, \qquad \sum_{j=1}^p\mid\beta_j\mid, \qquad \left(\sum_{j=1}^p\mid\beta_j\mid,\,\,\,\sum_{j=1}^p\beta_j^2\right)
\end{equation*}
respectively. This type of regularized WLS with lasso penalty is studied recently by Huang et al.~(2006\nocite{hua:ma:xie:06:regu}).
A regularized WLS method called CCLASSO where a combination of ridge and lasso penalty is used in Hu and Rao (2010\nocite{hu:rao:2010:sparse}).
So, the objective function of the regularized WLS method with censoring constraints becomes
\begin{align}
L(\beta,\, \lambda,\, \lambda_0)={}& \frac{1}{2}\,(Y_{u}-X_{u}\,\beta)^T (Y_{u}-X_{u}\,\beta) + \lambda\, \mbox{pen}(\beta)+\lambda_{0}\,\xi^T\xi,\nonumber\\
                                    &\text{subject to}~Y_{\bar{u}}\leq X_{\bar{u}}\,\beta+\xi.
\label{eq:regwlsecc}
\end{align}

\section{Proposed Model Framework}\label{sec:swlseReguNEW}
\subsection{The Regularized WLS: Adaptive Elastic Net (AEnet)}\label{sec:ch3.aenet}
The elastic net [Zou and Hastie (2005\nocite{zou:hast:05:regul})] has proved useful when analysing data with very many correlated covariates. The $\ell_1$ part of the penalty for elastic net generates a sparse model. On the other hand, the quadratic part of the penalty removes the limitation on the number of selected variables when $p \gg n$. The quadratic part of the penalty also stabilizes the $\ell_1$ regularization path and shrinks the coefficients of correlated predictors towards each other, allowing them to borrow strength from each other. The elastic net can not be applied directly to the AFT models because of censoring, but the regularized WLS (\ref{eq:regwlse}) with the elastic net penalty overcomes this problem. The naive elastic net estimator $\hat{\beta}$ for censored data is obtained as
\begin{equation}
\begin{array}{c}\arg\min\ \\ \beta \end{array}\frac{1}{2}\,(Y_{u}-X_{u}\,\beta)^T(Y_{u}-X_{u}\,\beta)+\lambda_1\,\sum_{j=1}^p|\beta_j|+\,\lambda_2\,\beta^T\beta.
\label{eq:enetSWLSE}
\end{equation}
With some algebra this naive elastic net can be transformed into a lasso type problem in an augmented space as below
\begin{equation*}
\begin{array}{c}\arg\min\ \\ \beta \end{array}\frac{1}{2}\,(Y^*_{u}-X^*_{u}\,\beta)^T(Y^*_{u}-X^*_{u}\,\beta)+\lambda_1\,\sum_{j=1}^p|\beta_j|,
\label{eq:enet2lassoSWLSE}
\end{equation*}
where
\begin{equation*}
X_{u}^*=\left(\begin{array}{c}X_{u} \\\sqrt{\lambda_2}\,I \end{array}\right)\qquad \mbox{and}\qquad Y_{u}^*=\left(\begin{array}{c}Y_{u} \\0 \end{array}\right).
\end{equation*}
The original elastic net estimator is now defined by
\begin{equation*}
\hat{\beta} (\mbox{elastic net})=(1+\lambda_2)\hat{\beta} (\mbox{naive elastic net}).
\end{equation*}

It is established in Zou (2006\nocite{Zou:06:theadaptive}) that the lasso does not exhibit the oracle properties. These properties include that the method selects the correct subset of predictors with probability tending to one, and estimates the non-zero parameters as efficiently as would be possible if we knew which variables were uninformative ahead of time. A modification, the adaptive elastic net, that does satisfy the oracle properties, was studied in Zou (2006\nocite{Zou:06:theadaptive}), Ghosh (2007\nocite{ghosh:07:adapEnet}), and Zou and Zhang (2009\nocite{zou:zha:09:OntheAdap}). The adaptive elastic net is a convex combination of the adaptive lasso penalty and the ridge penalty. Here we present an adaptive elastic net approach designed for censored data that is referred to as the AEnet approach throughout the paper. We introduce the adaptive elastic net penalty terms including coefficients to the regularized WLS objective function (\ref{eq:regwlse}).

\begin{equation}
\begin{array}{c}\arg\min\ \\ \beta \end{array}\frac{1}{2}\,(Y_{u}-X_{u}\,\beta)^T(Y_{u}-X_{u}\,\beta)+\lambda_1\,\sum_{j=1}^p\hat{w}_j\,|\beta_j|+\,\lambda_2\,\beta^T\beta,
\label{eq:aenetSWLSE}
\end{equation}
where $\hat{w}=1/|\hat{\beta_0}|^{\gamma}$ is the adaptive weight based on the initial estimator $\hat{\beta_0}$ for some $\gamma$. For the rest of this paper $w$ will denote weights obtained from an initial estimator. For the initial estimator $\hat{\beta_0}$, the OLS estimator  as suggested in Ghosh (2007\nocite{ghosh:07:adapEnet}) or the elastic net estimator as suggested in Zou and Zhang (2009\nocite{zou:zha:09:OntheAdap}) can be used. For this study we use $\gamma=1$ and the elastic net estimator on the weighted data as given by Equation (\ref{eq:enetSWLSE}) as the initial estimator $\hat{\beta_0}$.

The adaptive elastic net can be transformed into an adaptive lasso type problem in an augmented space in a similar way as for the naive elastic net.
\begin{equation}
\hat{\beta}_{a-nenet}^*=\begin{array}{c}\arg\min\ \\ \beta \end{array}\frac{1}{2}\,(Y^*_{u}-X^*_{u}\,\beta)^T(Y^*_{u}-X^*_{u}\,\beta)+\lambda_1\,\sum_{j=1}^p\hat{w}_j\,|\beta_j|,
\label{eq:anenet2alassoSWLSE}
\end{equation}
where
\begin{equation*}
{X_{u}^*}_{(n_u+p)\times p}=\left(\begin{array}{c}X_{u} \\\sqrt{\lambda_2}\,I \end{array}\right)\qquad \mbox{and}\qquad {Y_{u}^*}_{(n_u+p)\times p}=\left(\begin{array}{c}Y_{u} \\0 \end{array}\right).
\end{equation*}
So, for fixed $\lambda_2$ the adaptive elastic net is equivalent to an adaptive lasso in augmented space. The adaptive lasso estimates in (\ref{eq:anenet2alassoSWLSE}) can be solved for fixed $\lambda_2$ by the LARS algorithm [Efron et al.~(2004\nocite{efro:hast:john:tibs:04:least})]. According to \emph{Theorem 2} in Zou (2006\nocite{Zou:06:theadaptive}), and \emph{Theorem 3.2} in Ghosh (2007\nocite{ghosh:07:adapEnet}), the estimator $\hat{\beta}_{a-nenet}^*$ is asymptotically normal. Then the adaptive elastic net estimate can be obtained by rescaling the estimate found in Equation (\ref{eq:anenet2alassoSWLSE}).
\begin{equation*}
\hat{\beta}_{a-enet}^*=(1+\lambda_2)\,\hat{\beta}_{a-nenet}^*.
\end{equation*}

\subsubsection{AEnet Algorithm}
The algorithm for the proposed adaptive elastic net approach as shown below is referred to as the AEnet algorithm.
\\\\
\parbox{\linewidth}{
\textbf{Input:} Design matrix $X^{*}_{u}$, response $Y^{*}_{u}$, a fixed set for $\lambda_2$, and $\hat{w}$.
\begin{enumerate}
\item Define $X^{**}_{j(u)}=X^{*}_{j(u)}/\hat{w}_j, \,\,j=1,\cdots, p$.
\item Solve the lasso problem for all $\lambda_1$ and a fixed $\lambda_2$,\\
$\hat{\beta}_{a-nenet}^{**}=\begin{array}{c}\arg\min\ \\ \beta \end{array}\frac{1}{2}\,(Y^{*}_{u}-X^{**}_{u}\,\beta)^T(Y^{*}_{u}-X^{**}_{u}\,\beta)+\lambda_1\,\sum_{j=1}^p|\beta_j|.$
\item Calculate
$\hat{\beta_j}_{a-enet}^*=(1+\lambda_2)\,\hat{\beta_j}_{a-nenet}^{**}/\hat{w}_j.$
\end{enumerate}
}
\\

To find the optimal value for the tuning parameters $(\lambda_1,\,\lambda_2)$, $\lambda_2$ is typically assumed to take values in a relatively small grid, say (0,\, 0.5,\, 1.0,\, 1.5,\, 2.00,$\cdots$,5).
For each $\lambda_2$, the LARS algorithm produces the entire solution path. This gives the optimal equivalent specification for lasso in terms of fraction of the $\ell_1$ norm ($t_1$). Then the optimal pair of $(t_1,\,\lambda_2)$ is obtained using $k$-fold cross-validation.

\subsection{The Regularized WLS: Adaptive Elastic Net with Censoring Constraints (AEnetCC)}\label{sec:ch3.aenetcc}
Here we present an extension of the above adaptive elastic approach that allows the censoring constraints to be implemented into the optimization framework. The adaptive elastic net estimator for censored data given by (\ref{eq:aenetSWLSE}) can be rewritten with censoring constraints as
\begin{align}
\tilde{\beta}_{a-enet}^*={}&\begin{array}{c}\arg\min\ \\ \beta,\,\xi \end{array}\frac{1}{2}\,(Y_{u}-X_{u}\,\beta)^T(Y_{u}-X_{u}\,\beta)+\,\lambda_2\,\beta^T\beta+\lambda_{0}\,\xi^T\xi,\nonumber\\
                                    &\text{subject to}~\sum_{j=1}^p\hat{w}_j\,|\beta_j|\le t_1\,\,\text{and}\,\, Y_{\bar{u}}\leq X_{\bar{u}}\,\beta+\xi,
\label{eq:aenetccSWLSE}
\end{align}
where $t_1$ is the lasso tuning parameter. We use a quadratic programming (QP) approach to solve the minimization problem of (\ref{eq:aenetccSWLSE}).
%
The QP can not handle $|\beta_j|$ because the lasso constraint ($\sum_{j=1}^p|\beta_j|\le t_1$) makes the QP solutions nonlinear in the $Y_i$. Further modification is needed to use $|\beta_j|$ in the QP framework. Following Tibshirani (1996)\nocite{tibs:96:regres} we use a modified design matrix $\tilde{X}=[X,\,-X]$ and represent coefficients $\beta$ as the difference between two non-negative coefficients $\beta^+$ and $\beta^-$.
%
Although the technique doubles the number of variables in the problem, it requires only a (known and bounded) linear number of constraints, and only requires the solution to one QP problem. Now Equation (\ref{eq:aenetccSWLSE}) becomes
\begin{align}
\tilde{\beta}_{a-enet}^*={}&\begin{array}{c}\arg\min\ \\ \beta^+,\beta^-, \xi \end{array}\frac{1}{2}\,[Y_{u}-X_{u}\,(\beta^+-\beta^-)]^T[Y_{u}-X_{u}\,(\beta^+-\beta^-)]\nonumber\\
                                        &+\lambda_2\,\beta^{+^T}\beta^+
                                           +\lambda_2\,\beta^{-^T}\beta^-+\lambda_{0}\,\xi^T\xi,\nonumber\\
                                        &\text{subject to}~\sum_{j=1}^p\hat{w}_j\,(\beta_j^++\beta_j^-)\le t_1\nonumber\\
                                        &\text{and}~Y_{\bar{u}}\leq X_{\bar{u}}\,(\beta^+-\beta^-)+\xi,~\beta^+\ge 0,~\beta^-\ge 0.
\label{eq:aenetccSWLSE.qp}
\end{align}
According to Ghosh (2007\nocite{ghosh:07:adapEnet}), the estimator $\tilde{\beta}_{a-enet}^*$ is asymptotically normal.

\subsubsection{AEnetCC Algorithm}\label{ch3.ss.Alg.aenetcc}
The algorithm for the proposed adaptive elastic net with censoring constraints approach is referred to as the AEnetCC algorithm.\\\\
\parbox{\linewidth}{
\textbf{Input:} $\hat{w}$.
\begin{enumerate}
\item Define $X^{**}_{j(u)}=X^{}_{j(u)}/\hat{w}_j, \,\,j=1,\cdots, p$.
\item Solve the elastic net problem,
\begin{align}
\tilde{\beta}_{a-nenet}^{**}={}&\begin{array}{c}\arg\min\ \\ \beta^+, \beta^-, \xi \end{array}\frac{1}{2}\,[Y_{u}-X^{**}_{u}\,(\beta^+-\beta^-)]^T[Y_{u}-X^{**}_{u}\,(\beta^+-\beta^-)]\nonumber\\
                                        &+\lambda_2\,\beta^{+^T}\beta^+
                                           +\lambda_2\,\beta^{-^T}\beta^-+\lambda_{0}\,\xi^T\xi,\nonumber\\
                                        &\text{subject to}~\sum_{j=1}^p(\beta_j^++\beta_j^-)\le t_1\nonumber\\
                                        &\text{and}~Y_{\bar{u}}\leq X_{\bar{u}}\,(\beta^+-\beta^-)+\xi,~\beta^+\ge 0,~\beta^-\ge 0.\nonumber
\end{align}
\item Calculate
$\tilde{\beta_j}_{a-enet}^*=(1+\lambda_2)\,\tilde{\beta_j}_{a-nenet}^{**}/\hat{w}_j.$
\end{enumerate}
}\\

%
%
The AEnetCC has three tuning parameters $(\lambda_0,\,\lambda_1,\,\lambda_2)$. For this method we use the same optimal pair of $(\lambda_1,\,\lambda_2)$ as found in AEnet. Then $\lambda_0$ is typically allowed to take values in a grid such as (0,\, 0.5,\, 1,\, 1.5,$\cdots$,10), and the optimal value for $\lambda_0$ obtained by $5$-fold cross-validation. Here the value of $\lambda_0$ typically depends upon how stringently one wants the model to satisfy the censoring constraints compared to how good is the prediction for uncensored data.

\subsection{The Regularized WLS: Weighted Elastic Net (WEnet)}
In this section we present a \emph{weighted elastic net} for censored data. This is an extension of the adaptive elastic net where for suitable weight $w$ the ridge penalty term is expressed as $\sum_{j=1}^p(w_j\,\beta_j)^2,$ instead of $\sum_{j=1}^p\,\beta_j^2$. This is a doubly adaptive type model. This type of regularized technique for uncensored data was first studied in Hong and Zhang (2010\nocite{Hon:Zhang:2010:weighted}). They established the model consistency and its oracle property under some regularity conditions. Following the regularized WLS given in (\ref{eq:regwlse}) the weighted elastic net for censored data can be defined by

\begin{equation}
\begin{array}{c}\arg\min\ \\ \beta \end{array}\frac{1}{2}\,(Y_{u}-X_{u}\,\beta)^T(Y_{u}-X_{u}\,\beta)+n_{u}\,\lambda_1\,\sum_{j=1}^p\,w_j\,|\beta_j|+\frac {n_{u}}{2}\,\lambda_2\,\sum_{j=1}^p\,(w_j\,\beta_j)^2,
\label{eq:wenetSWLSE}
\end{equation}
where $w_j>0,\, j=1,\cdots,p$ are the weighted penalty coefficients. The weight is typically chosen as the standard deviations of the associated estimators [Hong and Zhang (2010\nocite{Hon:Zhang:2010:weighted})]. Since standard deviations are unknown in practice, we use the standard error of an initial consistent estimator. For estimating standard error under high--dimensional data we use a bootstrap procedure based on the elastic net model on the weighted data (\ref{eq:enetSWLSE}). For data where $n>p$, as in Jin et al.~(2003\nocite{jin:lin:wei:ying:03:rank}) and {Jin, Lin and Ying}~(2006{\nocite{jin:lin:ying:06:Onleast}}), we choose the Gehan type rank estimator as an initial estimator. This is defined [Gehan (1965\nocite{Gehan:65:Agen})] as the solution to the system of estimating equations, $0=U_G(\beta)$, where
\begin{equation}
U_G(\beta)=n^{-1} \sum_{i=1}^{n}\sum_{i'=1}^{n}\,\delta_i\, (X_i-X_{i'})\,I\{\xi_i(\beta) \leq \xi_{i'}(\beta)\},
\label{eq:Gehan2}
\end{equation}
and $\xi_i(\beta)=Y_i-X_i^T\,\beta$. Note that Equation (\ref{eq:Gehan2}) can be expressed as the $p$-dimensional gradient of the convex loss function, $n\,L_G(\beta)$, where
\begin{equation*}
L_G(\beta)= \sum_{i=1}^{n}\sum_{i'=1}^{n}\,\delta_i\{\xi_i(\beta)-\xi_{i'}(\beta)\}^-,
\end{equation*}
$a^-=\mathrm{1}_{\{a<0\}}|a|$.

Similarly to the adaptive elastic net the weighted elastic net can be transformed into a weighted lasso type problem on an augmented data set. We rewrite Equation (\ref{eq:wenetSWLSE}) with a scaled coefficient difference as
\begin{align}
\hat{\beta}_{w-enet}={}&\begin{array}{c}\arg\min\ \\ \beta \end{array}(Y_{u}-X_{u}\,\beta)^T(Y_{u}-X_{u}\,\beta)+\lambda_1\,\sum_{j=1}^p\,w_j\,|\beta_j|+\times\nonumber\\
                    &\lambda_2\,\sum_{j=1}^p\,(w_j\,\beta_j)^2\label{eq:wenetSWLSE2first}\\
                    =&\begin{array}{c}\arg\min\ \\ \beta \end{array}\left[\left(\begin{array}{c}Y_{u} \\0 \end{array}\right)-\left(\begin{array}{c}X_{u} \\ \sqrt{\lambda_2}\,W \end{array}\right)\frac{1}{\sqrt{1+\lambda_2}}\sqrt{1+\lambda_2}\,\beta\right]^T\times\nonumber\\
                    &\left[\left(\begin{array}{c}Y_{u} \\0 \end{array}\right)-\left(\begin{array}{c}X_{u} \\ \sqrt{\lambda_2}\,W \end{array}\right)\frac{1}{\sqrt{1+\lambda_2}}\sqrt{1+\lambda_2}\,\beta\right]+\frac{\lambda_1}{\sqrt{1+\lambda_2}}\,w_j\times\nonumber\\
                    &\sqrt{1+\lambda_2}\,|\beta_j|,
\label{eq:wenetSWLSE2}
\end{align}
where $W=\text{diag}[w_1,\cdots,w_p]$. Now assume that
\begin{align*}
{X_{u}^*}_{(n_u+p)\times p}={}&(1+\lambda_2)^{-\frac{1}{2}}\left(\begin{array}{c}X_{u} \\ \sqrt{\lambda_2}\,W \end{array}\right),\\
&{Y_{u}^*}_{(n_u+p)\times p}=\left(\begin{array}{c}Y_{u} \\0 \end{array}\right),\\
&\tilde{\lambda}=\frac{\lambda_1}{\sqrt{1+\lambda_2}},\\
&\beta^*=\sqrt{1+\lambda_2}\,\beta.
\end{align*}
Then the estimator in (\ref{eq:wenetSWLSE2}) with new notation becomes
\begin{align}
\hat{\beta}^*_{w-enet}=\begin{array}{c}\arg\min\ \\ \beta \end{array}(Y^*_{u}-X^*_{u}\,\beta^*)^T(Y^*_{u}-X^*_{u}\,\beta^*)+\tilde{\lambda}\sum_{j=1}^p\,w_j\,|\beta^*_j|\label{eq:wenetSWLSE3}.
\end{align}

So, for fixed $\lambda_2$, the weighted elastic net can be transformed into an adaptive lasso problem (\ref{eq:wenetSWLSE3}) in some augmented space. The weighted elastic net estimator $\hat{\beta}_{w-enet}$ can be obtained by
\begin{equation}
\hat{\beta_j}_{w-enet}=\hat{\beta_j}^*_{w-enet}/\sqrt{1+\lambda_2},
\label{eq:wenetSWLSEnaive}
\end{equation}
which is a naive elastic net estimator. The original weighted elastic net estimator $\hat{\beta_j}_{w-enet}$ is therefore obtained by
\begin{equation}
\sqrt{1+\lambda_2}\,\hat{\beta_j}^*_{w-enet}.
\label{eq:wenetSWLSEfinal}
\end{equation}

\subsubsection{WEnet Algorithm}
The algorithm for the proposed weighted elastic net approach is referred to as the WEnet algorithm.
\\\\
\parbox{\linewidth}{
\textbf{Input:} Design matrix $X^{*}_{u}$, response $Y^{*}_{u}$,\, $\tilde{\lambda}$, and $\hat{w}_j$ for $j=1,\cdots,p$.
\begin{enumerate}
\item Define $X^{**}_{j(u)}=X^{*}_{j(u)}/\hat{w}_j, \,\,j=1,\cdots, p$.
\item Solve the lasso problem for all $\lambda_1$ and a fixed $\lambda_2$,\\
$\hat{\beta}_{w-nenet}^{**}=\begin{array}{c}\arg\min\ \\ \beta \end{array}(Y^{*}_{u}-X^{**}_{u}\,\beta)^T(Y^{*}_{u}-X^{**}_{u}\,\beta)+\tilde{\lambda}\,\sum_{j=1}^p|\beta_j|.$
\item Calculate
$\hat{\beta}_{w-enet}^*=\hat{\beta}_{w-nenet}^{**}/\hat{w}_j.$
\end{enumerate}
\textbf{Output:} $\hat{\beta}_{w-enet}=\sqrt{(1+\lambda_2)}\,\hat{\beta}_{w-enet}^{*}.$
}\\

To find the optimal value for the tuning parameters $(\lambda_1,\,\lambda_2)$, $\lambda_2$ is typically assumed to take values in a relatively small grid, similar to the grid used for AEnet algorithm. To optimize the tuning parameters $(\lambda_1,\,\lambda_2)$ for WEnet we follow exactly the same procedure as described in the previous section for the AEnet.

\subsection{The Regularized WLS: Weighted Elastic Net with Censoring Constraints (WEnetCC)}
Following the adaptive elastic net with censoring constraints the weighted elastic net model with censoring constraints can be defined by
\begin{align*}
\tilde{\beta}_{w-enet}^*={}&\begin{array}{c}\arg\min\ \\ \beta,\,\xi \end{array}\frac{1}{2}\,(Y_{u}-X_{u}\,\beta)^T(Y_{u}-X_{u}\,\beta)+\lambda_2\,W\,\beta^{T}\beta+\lambda_{0}\,\xi^T\xi,\nonumber\\
                                    &\text{subject to}~\sum_{j=1}^p\hat{w}_j\,|\beta_j|\le t_1\,\,\text{and}\,\, Y_{\bar{u}}\leq X_{\bar{u}}\,\beta+\xi.
\label{eq:wenetSWLSEcc}
\end{align*}

Now after representing $\beta^*$ as the difference between two non-negative coefficients ${\beta^*}^+$ and ${\beta^*}^-$, the Equation (\ref{eq:wenetSWLSEcc})
becomes
\begin{eqnarray}
\tilde{\beta}^{**}_{w-enet}={}&\begin{array}{c}\arg\min\ \\ {\beta^*}^+,\, {\beta^*}^-,\, \xi \end{array}(Y^*_{u}-X^*_{u}\,({\beta^*}^+-{\beta^*}^-))^T(Y^*_{u}-X^*_{u}\,({\beta^*}^+-{\beta^*}^-))\nonumber\\
                                    &+\lambda_{0}\,\xi^T\xi,\nonumber\\
                                    &\text{subject to}~~\sum_{j=1}^p\,w_j\,({\beta^*}^+-{\beta^*}^-)\le t_1\,\,\text{and}\nonumber\\
                                    &Y_{\bar{u}}\leq X_{\bar{u}}\,({\beta^*}^+-{\beta^*}^-)+\xi.
\label{eq:wenetSWLSEccfinal}
\end{eqnarray}

\subsubsection{WEnetCC Algorithm}\label{ch3.ss.Alg.wenetcc}
Below we present the algorithm for the proposed weighted elastic net with censoring constraints approach which is referred to as the WEnetCC algorithm.
\\\\
\parbox{\linewidth}{
\textbf{Input:} Design matrix $X^{*}_{u}$, response $Y^{*}_{u}$,\, $\tilde{\lambda}$, and $\hat{w}_j$ for $j=1,\cdots,p$.
\begin{enumerate}
\item Define $X^{**}_{j(u)}=X^{*}_{j(u)}/\hat{w}_j, \,\,j=1,\cdots, p$.
\item Solve the lasso problem,
\begin{eqnarray*}
\tilde{\beta}^{**}_{w-enet}={}&\begin{array}{c}\arg\min\ \\ {\beta^*}^+,\, {\beta^*}^-,\, \xi \end{array}(Y^*_{u}-X^{**}_{u}({\beta^*}^+-{\beta^*}^-))^T(Y^*_{u}-X^{**}_{u}\nonumber\\
                                    &\times({\beta^*}^+-{\beta^*}^-))+\lambda_{0}\,\xi^T\xi,\nonumber\\
                                    &\text{subject to}~~\sum_{j=1}^p\,({\beta^*}^+-{\beta^*}^-)\le t_1\,\,\text{and}\nonumber\\
                                    &Y_{\bar{u}}\leq X_{\bar{u}}\,({\beta^*}^+-{\beta^*}^-)+\xi.
\label{eq:wenetSWLSEccfinal2}
\end{eqnarray*}
\item Calculate
$\tilde{\beta}_{w-enet}^*=\hat{\beta}_{w-enet}^{**}/\hat{w}_j.$
\end{enumerate}
\textbf{Output:} $\tilde{\beta}_{w-enet}=\sqrt{(1+\lambda_2)}\,\tilde{\beta}_{w-enet}^{*}.$
}\\

The estimator in the second step is obtained by optimizing the QP problem. 
%
%
To obtain optimal tuning parameter $(\lambda_0,\,\lambda_1,\,\lambda_2)$ of the WEnetCC we follow exactly the same procedure as
for the AEnetCC algorithm. According to \emph{Theorem 2} in Zou (2006\nocite{Zou:06:theadaptive}), and \emph{Theorem 3.2} in Ghosh (2007\nocite{ghosh:07:adapEnet}), both estimators $\hat{\beta_j}_{w-enet}$ and $\tilde{\beta_j}_{w-enet}$ are asymptotically normal.

\begin{remark}
Under some regularity conditions, the WLS estimator with K$-$M weights is consistent and asymptotically normal [Stute (1993, 1996\nocite{stut:93:consist}\nocite{stut:96:distrib})]. The proof is not directly applicable to the lasso penalty since the lasso penalty is not differentiable. In Huang {et al.}~(2006\nocite{hua:ma:xie:06:regu}) it is shown that the regularized WLS estimator with lasso penalty has the asymptotic normality property. In their proof they added two more conditions additional to the regularity conditions mentioned in Stute (1993, 1996\nocite{stut:93:consist}\nocite{stut:96:distrib}). The two additional conditions are (i) the regularized WLS lasso estimator has finite variance, and (ii) the bias of the K$-$M integrals is of the order $O(n^{1/2})$, which is related to the level of censoring and to the tail behavior of the K$-$M estimator.
\end{remark}

\subsection{Variable Selection Criteria}\label{ch3.sec:VSC}
For AEnet and WEnet we use LARS that produces exact zero coefficients in solution paths and hence does parsimonious variable selection. For the remaining two methods AEnetCC and WEnetCC we use an approach that simultaneously removes several variables that have very small coefficients.

%

We
use an $\mbox{AIC}_c$ type score based on the weighted $k$-fold cross-validation error CV$-$S (which is the sum of squared residuals of uncensored data multiplied by the K$-$M weights i.e.~$(Y_{u}-X_{u}\,\hat{\beta})^T \,w_u\,(Y_{u}-X_{u}\,\hat{\beta})$). The $\mbox{AIC}_c$ score is defined by
\begin{align}
\mbox{AIC}_c~\mbox{score} = n_{u}\log(\mbox{CV$-$S})+2k\Big(\frac{n_{u}}{n_{u}-k-1}\Big).
\end{align}

\subsubsection{Variable Selection Algorithm for AEnetCC and WEnetCC}
The above $\mbox{AIC}_c$ score is used as variable selection criteria.
\\\\
\parbox{\linewidth}{
\begin{enumerate}\itemsep1pt
    \item[1:]  Get the optimal pair ($\lambda_1,\,\lambda_2$) from fitting AEnet or WEnet.
    \item[2:] Fix a set of $\lambda_0$. Then for each $\lambda_0$
    \begin{enumerate}
    \item  Fit AFT model by the computational procedure (\ref{ch3.ss.Alg.aenetcc}) or (\ref{ch3.ss.Alg.wenetcc}). Find the predictor set PS by using $|\hat{\beta}|>\varsigma$.
    \item Use the PS and divide the dataset into $k$ parts. Leave out one part at a time and fit AFT model by the computational procedure (\ref{ch3.ss.Alg.aenetcc}) or (\ref{ch3.ss.Alg.wenetcc})\label{al:ch3.2b}.
    \item Combine the $K$--fitted models built in step 2(b) by averaging their coefficients. Compute the $\mbox{CV--S}$ and then $\mbox{AIC}_c$ score.
\end{enumerate}
    \item[3:]  Repeat steps 2 until all $\lambda_0$ are exhausted. Return the model with the lowest $\mbox{AIC}_c$ score and corresponding $\lambda_0$.
\end{enumerate}
}\\

We choose a very small value for precision parameter, say $\varsigma=1e^{-5}$, as a default value but any other suitable value can be chosen. Alternatively, $\varsigma$ should be considered as an additional tuning parameter in the above variable selection algorithm. It is also possible either to adapt existing efficient optimization algorithm such as SCAD [Fan and Li (2001\nocite{fan:li:01:varia})] or LARS-EN [a modified LARS developed for adaptive elastic net by Ghosh (2007\nocite{ghosh:07:adapEnet})] or to develop a new algorithm that avoids the $\varsigma$ parameter completely.

\subsection{Measures of Fit and Measures of Prediction}\label{sec4.msebootstrap}
The following MSE is computed to measure the fit in the training data:
\begin{align}
MSE_{TR} =\frac{1}{n_{u}} \sum_{i=1}^n\delta_i(\hat{Y}_i-Y_i)^2,
\label{eq:msef}
\end{align}
where $n_{u}$ is the number of uncensored observations. This measure compares the fitted values with the true values corresponding to the uncensored observations. We first generate a training dataset, such as $Y$ in (\ref{eq:msef}), and then a test dataset $Y_{\text{new}}$ (say) of the same size using the same design parameters. All the methods are fitted using the training data. Then in order to get predicted values $\hat{Y}_{\text{new}}$ the fitted model is used with the $\textbf{X}$ matrix of the test data. We measure the prediction accuracy by
\begin{align}
MSE_{TE} =\frac{1}{n} \sum_{i=1}^n(\hat{Y}_{\text{new},\,i}-Y_{\text{new},\,i})^2.
\label{eq:msep}
\end{align}

We estimate the variance of the regression parameters using the nonparametric 0.632 bootstrap [Efron and Tibshirani (1993)\nocite{efr:tib:93:book}] in which one samples $\tilde{n}\approx 0.632\times n$ from the $n$ observations without replacement. We use $\tilde{n}\approx 0.632\times n$ as the expected number of distinct bootstrap observations is about $0.632\times n$.
We use $B\geq 500$ and then the sample variance of the bootstrap estimates provide an estimate of the variance of $\hat{\beta}$.

\section{Simulation Studies}\label{sec3}
The purpose of this section is to evaluate and compare the performance of the proposed approaches using simulation studies and a real data example. We use six existing model selection approaches for comparison purpose. However the aim is not to address which approach is superior, rather we identify the similarities among the methods and accordingly provide some suggestions about the situations where one approach may outperform the others. The six approaches are used for the last simulation example and also for the real data example. For AEnet and WEnet methods we use \{0, 0.6, 1.1, 1.7, 2.2, 2.8, 3.3, 3.9, 4.4, 5.0\} as the grid for $\lambda_2$. For the two censoring constraint based methods AEnetCC and WEnetCC the set \{0, 1.0, 1.4, 1.8, 2.2, 2.6, 3.0\} is used for the penalties of violations of constraints $\lambda_0$.

Several alternative penalized regression and Bayesian approaches for variable selection for high--dimensional censored data have been developed. We use the simple elastic net (Enet) approach as defined in Equation (\ref{eq:enetSWLSE}) on the weighted data.
Another approach is the adaptive elastic net for AFT (ENet-AFT) in Engler and Li~(2009\nocite{eng:li:09:surv}). There is also an MCMC selection based Bayesian method (Bayesian-AFT) for log-normal AFT model introduced in Sha, Tadesse and Vannucci~(2006\nocite{sha:tad:van:06:bay}); we use this approach only for the log-normal AFT model. We also use three correlation-based greedy variable selection approaches: sure independence screening (SIS) [Fan and Lv (2008\nocite{Fan:Lv:08:Sure})], tilted correlation screening (TCS) [Cho and Fryzlewicz (2012\nocite{Cho:Fryz:12:HighDim})], and PC-simple [B\"{u}hlmann, Kalisch and Maathuiset (2010\nocite{Buh:kal:maat:10:varia})] all implemented with the weighted data under the SWLS as defined by the Equation (\ref{eq:swlse}).

\subsection{Simulation Studies}\label{ch3.sec.simu}
The logarithm of the survival time is generated from the true AFT model
\begin{equation}
Y_i=\alpha+X_i^T\beta +\sigma \varepsilon_i,~~i=1,\cdots,n
\label{eq:ch3.aft}
\end{equation}
with $\varepsilon_i\sim f(\cdot)$, any suitable probability density function and $\sigma$, the signal to noise ratio. We use correlated datasets.
Censoring time is generated from particular distributions maintaining a desired censoring level, $P_{\%}$. We consider three $P_{\%}$: 30, 50, and 70 that are indicated as low, medium and high respectively. We maintain random censoring except for the case that if the largest observation is found censored $(i.e.~Y^+_{(n)})$ then we reclassify it as uncensored according to Efron's tail correction [Efron (1967\nocite{efro:67:the})]. This is necessary since the WLS method involves the K$-$M weights that are based on the K$-$M distribution function [Khan and Shaw (2013\nocite{Kha:sha:13:OnDeal})].

\subsubsection{Simulation I: $n=100$, $p=40$}\label{sec:simIIa}
We consider $40$ covariates with three blocks: $\beta$ coefficients for $j\in\,\{1,\cdots,5\}$ are set to be 5 and for $j\in\,\{6,\cdots,10\}$ are set to be 2. We treat these two blocks as informative blocks i.e.~contain potential covariates (say, $p_{\gamma}$ = 10). The remaining $\beta$ coefficients (i.e.~$j\in\,\{11,\cdots,40\}$) are set to be zero and we set $\mathbf{X}\sim U(0,1)$.
We consider log-normal and Weibull AFT models. The survival time is generated using (\ref{eq:ch3.aft}) with $\varepsilon_i\sim N(0,1)$ for the log-normal AFT model and using (\ref{eq:ch3.aft}) with $\varepsilon_i\sim \mbox{log(Weibull)}$ for the Weibull AFT model. More precisely, in Equation (\ref{eq:ch3.aft}) the error is $\sigma \varepsilon$ where $\sigma=1$ and
\begin{equation*}
\varepsilon = \frac{\log[W]-E \log [W]}{\sqrt{\mbox{Var}\{\log[W]\}}},~~~ \mbox{where}\,\,W\sim \mbox{Weibull (5,\,1).}
\end{equation*}
For both AFT models the censoring time is generated using the log-normal distribution $\exp[N(c_0\sqrt{1+\sigma},~(1+\sigma^2))]$. Here $c_0$ is calculated analytically to produce the chosen $P_{\%}$. We fit all four methods and for both AFT models 100 runs are simulated. For each covariate we record the frequency of being selected among 100 simulation runs, the minimum, mean and maximum. The summary statistics for each block are shown in Table \ref{tab:simex2}.

\begin{table}[ht]\centering
\caption{Variable selection frequency percentages for the methods for both log-normal and Weibull AFT models. Here (*) stands for the noninformative block with zero coefficients.}
\scalebox{0.6}{
\begin{tabular}{lllcccc}\toprule\toprule
$P_{\%}$&Methods&Parameters&\multicolumn{2}{c}{$r_{ij}$ = 0}&\multicolumn{2}{c}{$r_{ij}$ = 0.5}\\
&&&log-normal&Weibull&log-normal&Weibull\\
&&&(Min,~Mean,~Max)&(Min,~Mean,~Max)&(Min,~Mean,~Max)&(Min,~Mean,~Max)\\\midrule 
\multirow{12}{*}{30}&\multirow{3}{*}{AEnet}& Block1&(99,~99,~99)&(95,~95,~95)&(62,~69.6,~79)&(62,~66.6,~74)\\
&&Block2&(66,~74.6,~83)&(73,~94.8,~97)&(39,~43.8,~51)&(35,~41.8,~48)\\
&&Block3*&(00,~1.6,~05)&(00,~2.8,~07)&(11,~18.5,~25)&(12,~18.9,~24)\\
&\multirow{3}{*}{AEnetCC}
&Block1&(100,~100,~100)&(100,~100,~100)&(76,~80,~82)&(80,~87,~95)\\
&&Block2&(48,~53.6,~57)&(48,~51.8,~54)&(44,~46.8,~50)&(51,~54.4,~59)\\
&&Block3*&(00,~3.1,~06)&(02,~6.3,~12)&(17,~25.8,~33)&(22,~30,~38)\\
&\multirow{3}{*}{WEnet}
&Block1&(99,~99,~99)&(99,~99,~99)&(100,~100,~100)&(100,~100,~100)\\
&&Block2&(63,~73.4,~82)&(74,~77.2,~79)&(51,~62.8,~75)&(55,~62.2,~68)\\
&&Block3*&(00,~1.3,~3)&(00,~1.9,~06)&(00,~2.3,~06)&(00,~03,~07)\\
&\multirow{3}{*}{WEnetCC}
&Block1&(100,~100,~100)&(99,~99.8,~100)&(93,~95.4,~97)&(96,~97,~98)\\
&&Block2&(30,~36,~39)&(24,~32.4,~36)&(44,~47.6,~52)&(43,~49,~57)\\
&&Block3*&(00,~2.2,~5)&(00,~2.5,~06)&(04,~9.6,~15)&(03,~9.2,~14)\\\\
\multirow{12}{*}{50}&\multirow{3}{*}{AEnet}&
Block1&(99,~99.6,~100)&(98,~98.8,~99)&(77,~81.8,~85)&(77,~81.6,~84)\\
&&Block2&(63,~69.8,~75)&(61,~66.6,~76)&(48,~52.2,~55)&(48,~53.6,~59)\\
&&Block3*&(09,~18.2,~25)&(10,~19,~27)&(22,~31.5,~38)&(26,~31.7,~41)\\
&\multirow{3}{*}{AEnetCC}&
Block1&(97,~98.6,~100)&(98,~99.2,~100)&(53,~56.4,~59)&(52,~57.8,~65)\\
&&Block2&(27,~33.6,~39)&(34,~38.2,~41)&(19,~22.6,~28)&(16,~24,~29)\\
&&Block3*&(00,~3.5,~08)&(01,~4.4,~10)&(05,~10.5,~18)&(04,~8.7,~13)\\
&\multirow{3}{*}{WEnet}&
Block1&(99,~99.8,~100)&(100,~100,~100)&(98,~99,~100)&(99,~99.6,~100)\\
&&Block2&(63,~68.6,~74)&(62,~65.8,~71)&(55,~59.4,~62)&(46,~58.2,~68)\\
&&Block3*&(05,~9.9,~18)&(05,~9.1,~14)&(01,~5.3,~10)&(02,~4.9,~09)\\
&\multirow{3}{*}{WEnetCC}&
Block1&(91,~95.8,~100)&(91,~95.4,~98)&(71,~76.2,~81)&(71,~74.2,~81)\\
&&Block2&(15,~21.6,~26)&(22,~23.4,~24)&(20,~25.4,~29)&(20,~26,~29)\\
&&Block3*&(00,~2.8,~7)&(00,~2.8,~07)&(02,~6.6,~12)&(02,~6.8,~12)\\\\
\multirow{12}{*}{70}&\multirow{3}{*}{AEnet}&
Block1&(96,~97.4,~99)&(93,~94.8,~97)&(71,~77,~82)&(70,~73.8,~77)\\
&&Block2&(52,~59.2,~62)&(56,~58.2,~62)&(52,~55,~58)&(42,~48.6,~55)\\
&&Block3*&(15,~23.5,~31)&(17,~24.1,~34)&(27,~34.9,~43)&(21,~28.7,~36)\\
&\multirow{3}{*}{AEnetCC}&
Block1&(85,~89.8,~94)&(85,~88.6,~92)&(35,~37.2,~41)&(36,~41.6,~48)\\
&&Block2&(19,~27,~33)&(24,~29.4,~36)&(16,~17,~18)&(13,~19.8,~26)\\
&&Block3*&(03,~10,~15)&(05,~9.5,~14)&(05,~8.4,~12)&(05,~10,~16)\\
&\multirow{3}{*}{WEnet}&
Block1&(96,~97.2,~99)&(92,~96,~99)&(90,~93.8,~96)&(93,~94.6,~97)\\
&&Block2&(41,~49.4,~57)&(42,~50.2,~55)&(36,~44.8,~52)&(43,~49.2,~57)\\
&&Block3*&(07,~16.9,~22)&(13,~17.8,~24)&(08,~12.3,~18)&(05,~10.6,~18)\\
&\multirow{3}{*}{WEnetCC}&
Block1&(82,~85.8,~89)&(83,~86,~89)&(50,~57.8,~63)&(47,~57.6,~65)\\
&&Block2&(14,~23.4,~31)&(21,~25.2,~31)&(15,~19.6,~24)&(19,~22.2,~28)\\
&&Block3*&(03,~8.8,~15)&(03,~6.3,~12)&(03,~7.3,~11)&(02,~7.4,~15)\\
\bottomrule\bottomrule
\end{tabular}}
\label{tab:simex2}
\end{table}

As expected the covariates of the informatics blocks (blocks 1 and 2) should be selected. For both AFT models Table \ref{tab:simex2} shows that when the data is uncorrelated all the methods tend to select most of the informative covariates from block 1 and with a very high percentage of informative covariates from block 2. The inclusion rate of the noninformative covariates into the final models of the methods is shown to be very low particularly with low censoring. When the covariates are correlated the two methods WEnet and WEnetCC outperform the other two methods. The methods AEnet and AEnetCC tend to select informative covariates with low selection frequencies, and tend to include more irrelevant predictors in their models. However, as censoring increases the overall performance in terms of both selecting informative covariates and excluding noninformative covariates slightly decreases.

\begin{figure}[ht]
\centering
\includegraphics[scale=0.4]{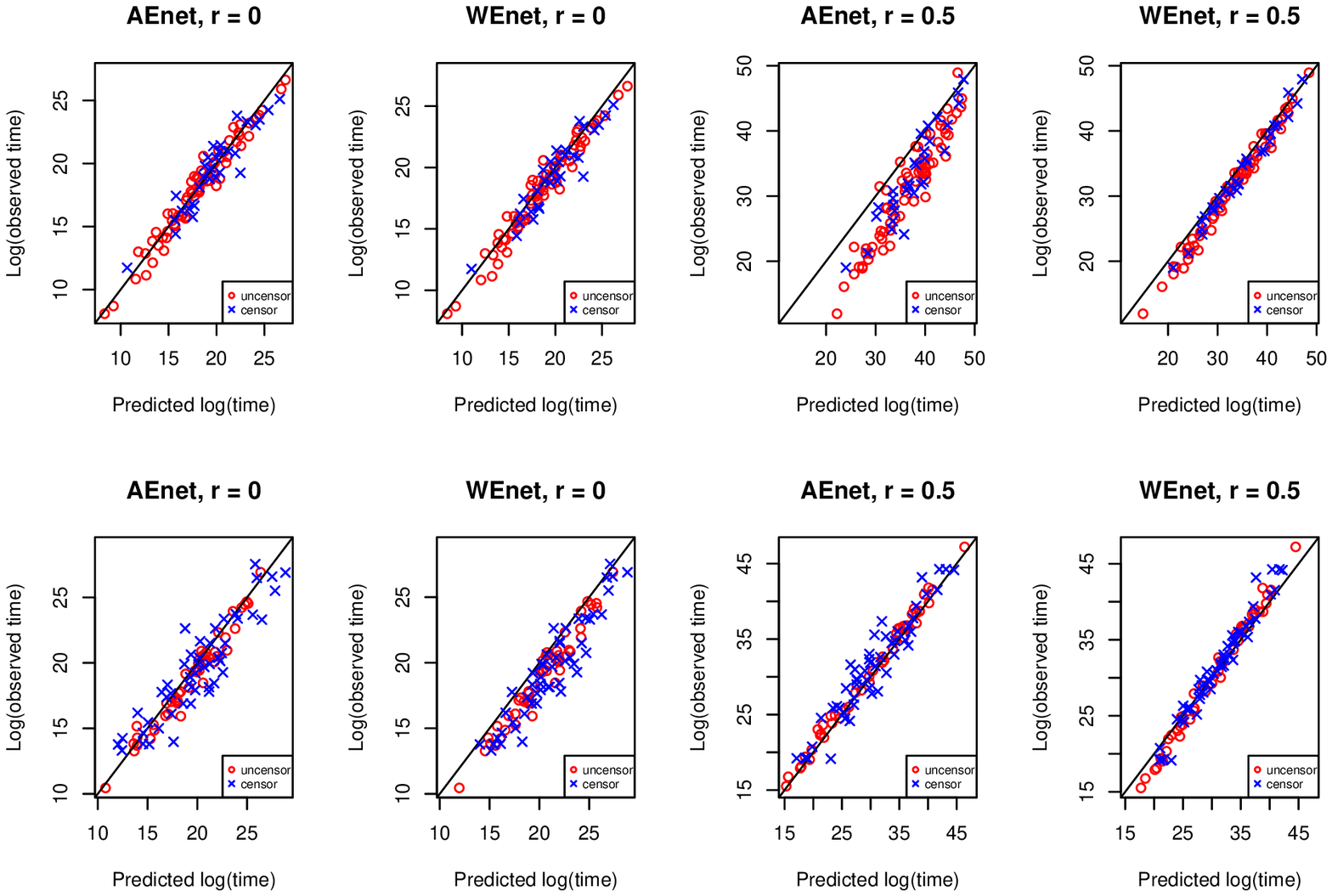}\\
\includegraphics[scale=0.4]{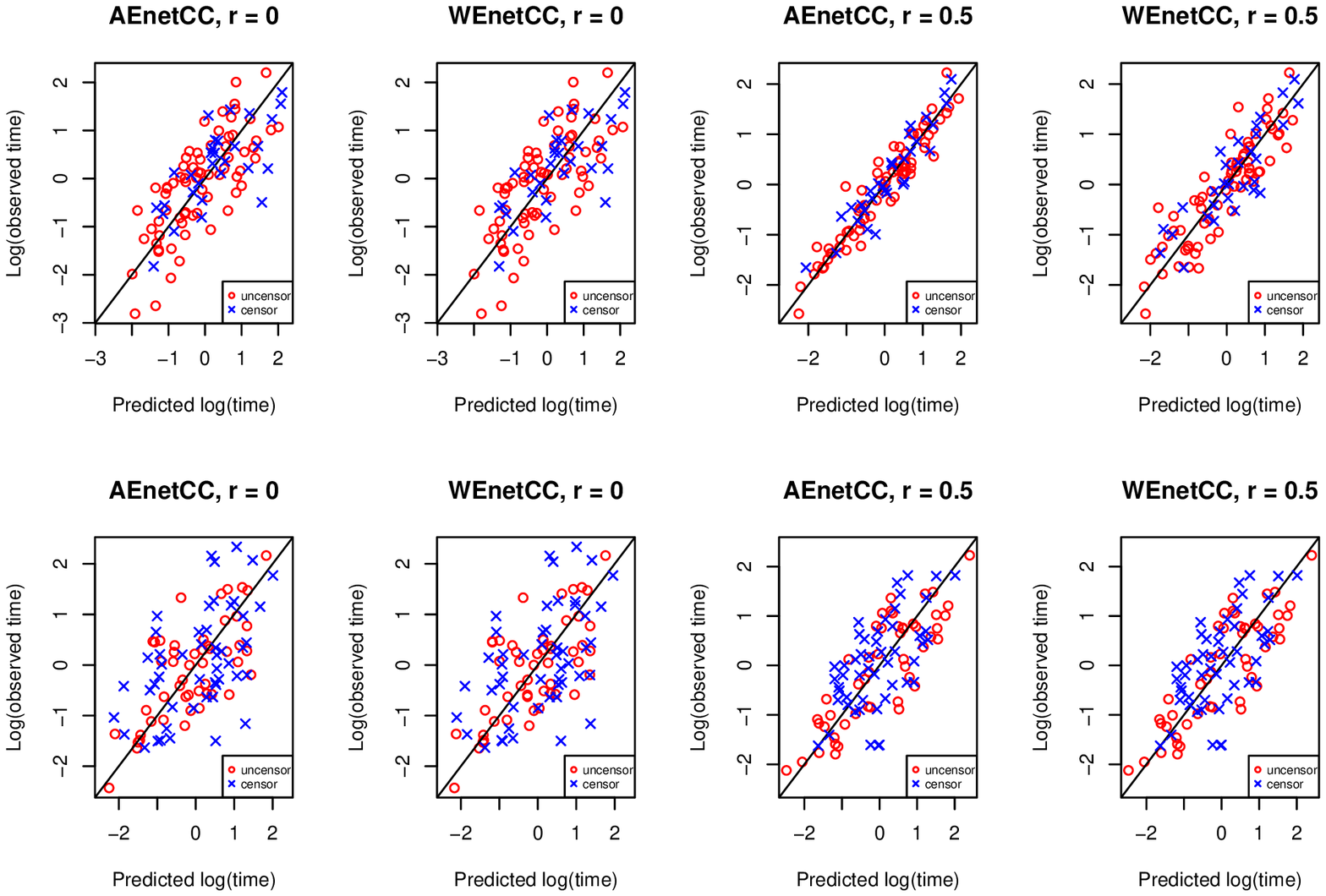}
\caption{Predicted vs observed log survival time under log-normal AFT model for the methods AEnet and WEnet for datasets with $P_{\%}= 30$ (first row panel) and $P_{\%}= 50$ (second row panel) and for the methods AEnetCC and WEnetCC for datasets with $P_{\%}= 30$ (third row panel) and $P_{\%}= 50$ (fourth row panel)}.
\label{fig:simex2.ln.3050}
\end{figure}

%
The prediction performance of the methods on the datasets under the log-normal AFT model is given in Figures \ref{fig:simex2.ln.3050}
We similarly obtained graphs for the Weibull AFT model and not reported here since we got almost similar results as for the log-normal model.
According to those figures all the methods fit the test data reasonably well for all levels of censoring. For the AEnet on the correlated dataset with 30\% censoring, the predicted times are considerably biased compared the corresponding observed times (Figure \ref{fig:simex2.ln.3050}, row 1, column 3). However, this does not happen with 50\% censoring (row 2, column 3). So it is hard to draw general conclusion. Given this performance of AEnet, all other methods select the correct set of nonzero coefficients, although the coefficients may be poorly estimated.

\subsubsection{Simulation II: $n=100$, $p=120$}\label{sec:simIII}
We fix $\alpha=1$ and set the first 20 coefficients for $\beta$'s to 4 (i.e.~$p_{\gamma}$ is 20) and the remaining coefficients of $\beta$ to zero. We keep everything else similar to simulation I.
We fit all four proposed methods together with six other existing methods and compare them in terms of
variable selection. The Bayesian-AFT MCMC sampler was run for 200,000 iterations at which the first 100,000 iterations are used as burn-in. A starting model with 40 randomly selected variables is considered. For the Bayesian-AFT, we choose $0.01$ to be the cut--off for the marginal posterior probabilities. 
%

\begin{table}[ht]\centering
\caption{Variable selection frequency percentages for 20 $p_{\gamma}$ variables and 100 $p-p_{\gamma}$ (non-relevant) variables with the methods for both log-normal and Weibull AFT models.}
\scalebox{0.60}{
\begin{tabular}{lllcccc}\toprule\toprule
$P_{\%}$&Methods&Parameters&\multicolumn{2}{c}{$r_{ij}$ = 0}&\multicolumn{2}{c}{$r_{ij}$ = 0.5}\\
&&&log-normal&Weibull&log-normal&Weibull\\
&&&(Min,~Mean,~Max)&(Min,~Mean,~Max)&(Min,~Mean,~Max)&(Min,~Mean,~Max)\\\midrule 
\multirow{20}{*}{30}&\multirow{2}{*}{AEnet}& $p_{\gamma}$&(77,~84.8,~94)&(82,~87.9,~94)&(42,~50.7,~58)&(44,~54.2,~63)\\
&&$p-p_{\gamma}$&(02,~9.4,~19)&(02,~9.8,~17)&(11,~21.1,~29)&(17,~23.8,~31)\\
&\multirow{2}{*}{AEnetCC}
&$p_{\gamma}$&(77,~89.8,~98)&(83,~89.5,~94)&(49,~62,~70)&(60,~67,~77)\\
&&$p-p_{\gamma}$&(08,~14.3,~21)&(04,~13.3,~21)&(30,~40.3,~52)&(35,~44.67,~53)\\
&\multirow{2}{*}{WEnet}
&$p_{\gamma}$&(72,~80,~90)&(73,~80,~89)&(36,~43.3,~50)&(36,~43.2,~55)\\
&&$p-p_{\gamma}$&(05,~12.3,~21)&(05,~13.1,~22)&(00,~1.4,~05)&(00,~1.5,~05)\\
&\multirow{2}{*}{WEnetCC}
&$p_{\gamma}$&(82,~87.3,~94)&(82,~86.5,~91)&(52,~61.8,~73)&(55,~65,~75)\\
&&$p-p_{\gamma}$&(06,~12.4,~19)&(03,~11.6,~20)&(03,~11,~22)&(03,~10.5,~21)\\
&\multirow{2}{*}{Enet}
&$p_{\gamma}$&(96,~97.2,~98)&(97,~97.3,~98)&(96,~98.2,~100)&(96,~98.8,~100)\\
&&$p-p_{\gamma}$&(05,~6.5,~8)&(4,~6.5,~9)&(01,~4.3,~10)&(00,~4.1,~10)\\
&\multirow{2}{*}{Enet-AFT}
&$p_{\gamma}$&(50,~53.2,~56)&(50,~53.3,~55)&(36,~40.3,~48)&(32,~39.5,~50)\\
&&$p-p_{\gamma}$&(03,~4.3,~06)&(03,~4.4,~06)&(05,~16.7,~27)&(08,~16.7,~27)\\
&\multirow{2}{*}{Bayesian-AFT}
&$p_{\gamma}$&(45,~64,~75)&-&(45,~72,~85)&-\\
&&$p-p_{\gamma}$&(14,~29,~41)&-&(36,~48.3,~62)&-\\
&\multirow{2}{*}{SIS}
&$p_{\gamma}$&(11,~20.5,~29)&(13,~21.3,~29)&(06,~11.9,~18)&(06,~11.5,~18)\\
&&$p-p_{\gamma}$&(00,~0.9,~03)&(00,~0.8,~04)&(00,~2.6,~06)&(00,~2.7,~07)\\
&\multirow{2}{*}{TCS}
&$p_{\gamma}$&(38,~45.9,~53)&(44,~47.4,~52)&(42,~52.8,~61)&(44,~55.1,~64)\\
&&$p-p_{\gamma}$&(00,~3.4,~08)&(00,~3.7,~09)&(05,~15.1,~26)&(08,~15.0,~27)\\
&\multirow{2}{*}{PC-simple}
&$p_{\gamma}$&(11,~23.5,~30)&(16,~25.2,~33)&(11,~18.1,~26)&(13,~19.7,~29)\\
&&$p-p_{\gamma}$&(00,~1.1,~04)&(00,~0.9,~05)&(00,~5.6,~13)&(01,~5.4,~15)\\\\
\multirow{20}{*}{50}&\multirow{2}{*}{AEnet}&
$p_{\gamma}$&(59,~68.1,~75)&(60,~66.2,~73)&(41,~50.7,~56)&(43,~50.6,~60)\\
&&$p-p_{\gamma}$&(03,~8.8,~16)&(03,~9.7,~18)&(16,~26,~35)&(15,~25.6,~38)\\
&\multirow{2}{*}{AEnetCC}&
$p_{\gamma}$&(69,~76.4,~85)&(68,~72.7,~80)&(47,~57.2,~67)&(41,~52.5,~64)\\
&&$p-p_{\gamma}$&(16,~23.6,~31)&(12,~22.2,~33)&(26,~37.9,~50)&(25,~35.2,~50)\\
&\multirow{2}{*}{WEnet}
&$p_{\gamma}$&(45,~56.8,~65)&(46,~52.2,~60)&(21,~28.4,~36)&(21,~28.6,~37)\\
&&$p-p_{\gamma}$&(04,~10.6,~18)&(03,~11.5,~20)&(00,~1.8,~06)&(00,~02,~06)\\
&\multirow{2}{*}{WEnetCC}& $p_{\gamma}$&(69,~75.4,~87)&(68,~72.8,~80)&(49,~55.5,~66)&(46,~55.8,~64)\\
&&$p-p_{\gamma}$&(12,~21.1,~29)&(11,~21.2,~30)&(06,~12.4,~19)&(03,~12.3,~21)\\
&\multirow{2}{*}{Enet}
&$p_{\gamma}$&(64,~67.4,~70)&(66,~68.1,~70)&(70,~76.2,~85)&(71,~80.4,~85)\\
&&$p-p_{\gamma}$&(11,~12.7,~15)&(10,~12.7,~15)&(03,~10.3,~18)&(03,~8.8,~16)\\
&\multirow{2}{*}{Enet-AFT}
&$p_{\gamma}$&(33,~35.3,~38)&(33,~35.8,~38)&(25,~17.8,~41)&(26,~33.6,~41)\\
&&$p-p_{\gamma}$&(03,~04,~05)&(03,~04,~06)&(09,~17.8,~27)&(08,~17.9,~26)\\
&\multirow{2}{*}{Bayesian-AFT}
&$p_{\gamma}$&(40,~53.5,~70)&-&(30,~51.5,~60)&-\\
&&$p-p_{\gamma}$&(27,~50,~67)&-&(32,~45.1,~56)&-\\
&\multirow{2}{*}{SIS}
&$p_{\gamma}$&(11,~17.4,~25)&(09,~16.8,~23)&(05,~10.5,~15)&(06,~09.9,~17)\\
&&$p-p_{\gamma}$&(00,~1.5,~06)&(00,~1.6,~05)&(00,~2.9,~08)&(00,~3.0,~09)\\
&\multirow{2}{*}{TCS}
&$p_{\gamma}$&(48,~56.9,~64)&(46,~55.6,~64)&(40,~47.5,~55)&(38,~45.5,~59)\\
&&$p-p_{\gamma}$&(25,~37.6,~48)&(27,~37.9,~47)&(28,~39.5,~53)&(28,~39.9,~51)\\
&\multirow{2}{*}{PC-simple}
&$p_{\gamma}$&(13,~19.4,~27)&(12,~18.8,~26)&(07,~14.8,~21)&(10,~15.9,~23)\\
&&$p-p_{\gamma}$&(00,~1.9,~07)&(00,~1.9,~05)&(00,~5.3,~16)&(00,~5.1,~13)\\
\bottomrule\bottomrule
\end{tabular}}
\label{tab:simex3b}
\end{table}
Table \ref{tab:simex3b} shows the results from 100 simulation runs.
We evaluate the frequency of being selected among 100 simulation and then compute the minimum, mean, and maximum of those frequencies. The results are presented in the table for two censoring level, 30\% and 50\%.
In terms of the mean selection frequencies of informative variables, all proposed methods outperform the Enet-AFT, SIS, TCS, and PC-simple methods. Their performances are very close to the performance of the Enet method at the higher censoring level, although they show slightly poorer performances with lower censoring. However for the uncorrelated dataset and both AFT models, the two methods AEnet and AEnetCC tend to exclude fewer noninformative covariates. The three greedy approaches tend to select far fewer variables (both informative and spurious) in the final model. As expected, for all, approaches the number of variables selected from $p_{\gamma}$ decreases as the censoring increases.
%
%
%
%
\section{Real Data Example}\label{sec4}
\subsection{Mantle cell lymphoma data}\label{sec:mcldata}
\begin{figure}[ht]
\centering
\includegraphics[scale=0.50]{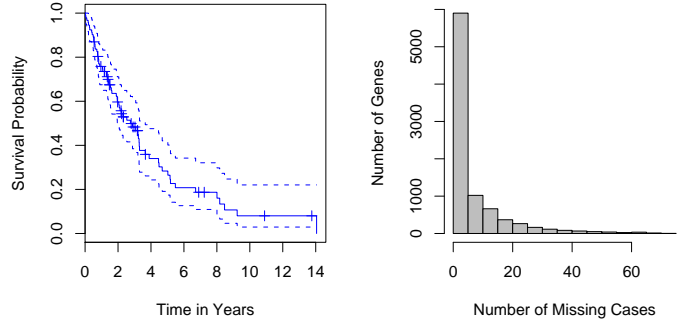}
\caption{(a) K--M plots of overall survival of patients for MCL data (left panel). (b) Histogram of number of genes with missingness (right panel).}
\label{fig:mclsummary}
\end{figure}

Rosenwald et al.~(2003)\nocite{ros:wri:wie:cha:etal:03:the} reported a study using microarray expression analysis of mantle cell lymphoma (MCL). The primary goal of the study was to discover gene expression signatures that correlate with survival in MCL patients. MCL accounts for 6\% of all non$-$Hodgkins lymphomas and a higher fraction of deaths from lymphoma, given that it is an incurable malignancy [Swerdlow and Williams (2002\nocite{swe:wil:2002:canthe})]. Among 101 untreated patients with no history of previous lymphoma included in the study, 92 were classified as having MCL, based on established morphologic and immunophenotypic criteria. Survival times of 64 patients were available and the remaining 28 patients were censored (i.e.~censoring rate $P_{\%}$ is $30$). The median survival time was 2.8 years (range 0.02 to 14.05 years). The length of survival of MCL patients following diagnosis is quite heterogeneous (Figure
\ref{fig:mclsummary} (a)). Many patients died within the first 2 years following diagnosis, yet 15\% (14/92) of the patients survived more than 5 years and 3 patients survived more than 10 years.
Lymphochip DNA microarrays were used to quantify mRNA expression in the lymphoma samples from the 92 patients. The gene expression dataset that contains expression values of 8810 cDNA elements is available at http://llmpp.nih.gov/MCL/. The data do not provide any further relevant covariates for MCL patients.
%

We apply the AFT model with all methods to this dataset. Although these methods have in principle no limit to the number of genes that can be used, we pre-process the data in a simple way. Pre-processing is important to gain further stability by reducing potential noise from the dataset. The pre-processing steps can be summarized as follows: (1) First, missing values of the original dataset are imputed by their sample means (for example, for a particular gene the missing gene expression value for a patient is replaced by the mean of the gene expression values for the observed patients). (2) Secondly, we compute correlation coefficients of the uncensored survival times with gene expressions. (3) Finally, a reasonable number of genes are selected based on their correlation with the response. After pre-processing, 574 genes with the largest absolute correlation ($>$0.3) have been identified and selected for analysis.
We then standardize these 574 gene expressions to have zero mean and unit variance and take logarithms of the observed times.
We use mean imputation to impute missing values for the MCL dataset.

We employ all the approaches and select the optimal tuning parameter with 5-fold cross-validation. The results are reported in Table \ref{tab:mclcompsummary}. The results suggest that most of the methods select a considerable number of genes and there are many common genes that are found between the methods. The three greedy methods tend to select fewer genes. The TCS selects the lowest number of genes (2) while the Enet selects the largest number of genes (68). Among the four proposed methods the AEnetCC selects the lowest number of genes (18).
The Enet, Enet-AFT, Bayesian-AFT, SIS, TCS, and PC-simple select 3, 5, 4, 5, 1, 1 genes respectively.

\begin{table}[ht]
\centering \caption{Number of genes selected by the methods (diagonal elements) and number of common genes found between the methods (off diagonal elements).}
\scalebox{0.6}{\begin{tabular}{lcccccccccc}\toprule\toprule
Methods&AEnet&AEnetCC&WEnet&WEnetCC&Enet&Enet-AFT&Bayesian-AFT&SIS&TCS&PC-simple\\\hline
AEnet&\color{red}{45}&12&10&03&10&07&05&05&01&02 \\
AEnetCC&12&\color{red}{18}&02&01&02&05&04&05&01&01 \\
WEnet&10&02&\color{red}{39}&09&03&01&01&00&00&00\\
WEnetCC&03&01&09&\color{red}{40}&02&01&02&01&00&00\\
Enet&10&02&03&02&\color{red}{68}&08&03&01&00&01\\
Enet-AFT&07&05&01&01&08&\color{red}{25}&05&03&01&00 \\
Bayesian-AFT&05&04&01&02&03&05&\color{red}{25}&02&01&00\\
SIS&05&05&00&01&01&03&02&\color{red}{05}&01&01\\
TCS&01&01&00&00&00&01&01&01&\color{red}{02}&00\\
PC-simple&02&01&00&00&01&00&00&01&00&\color{red}{03}\\
\bottomrule\bottomrule
\end{tabular}}
\label{tab:mclcompsummary}
\end{table}

\begin{figure}[ht]
\centering
\includegraphics [height=3.5cm, width=4.5cm] {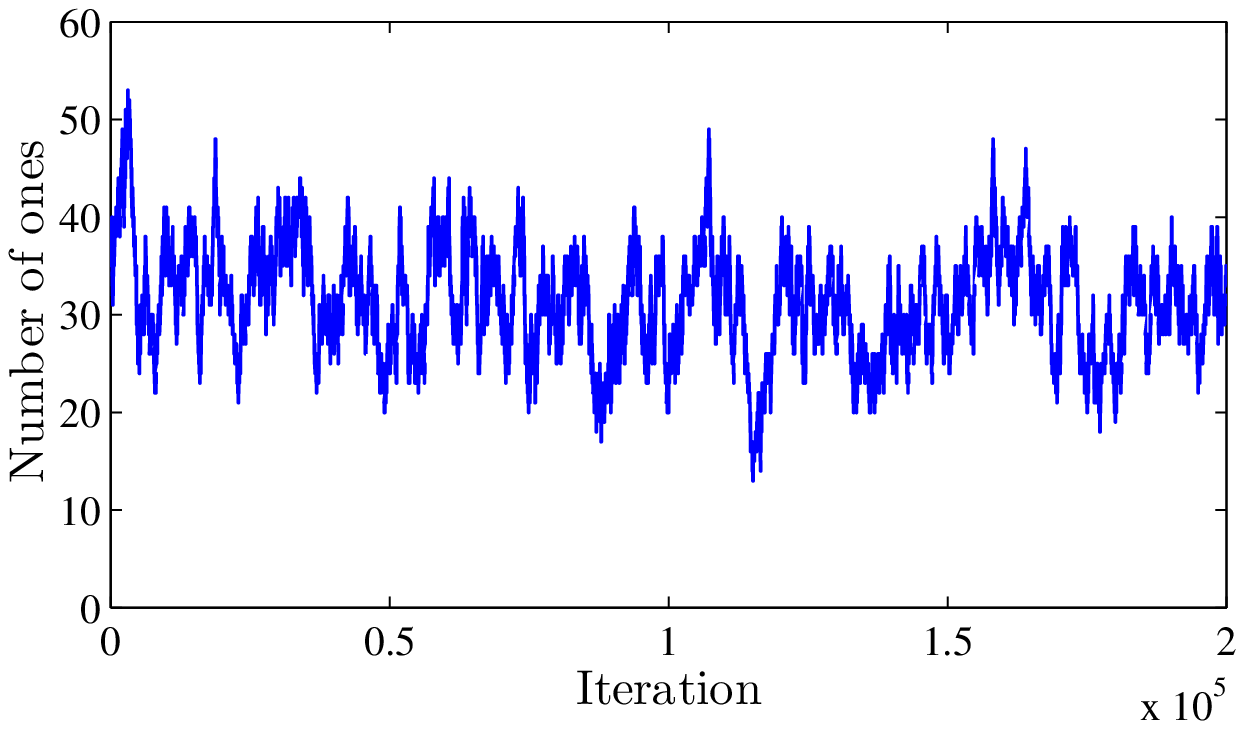}\includegraphics [height=3.5cm, width=4.5cm] {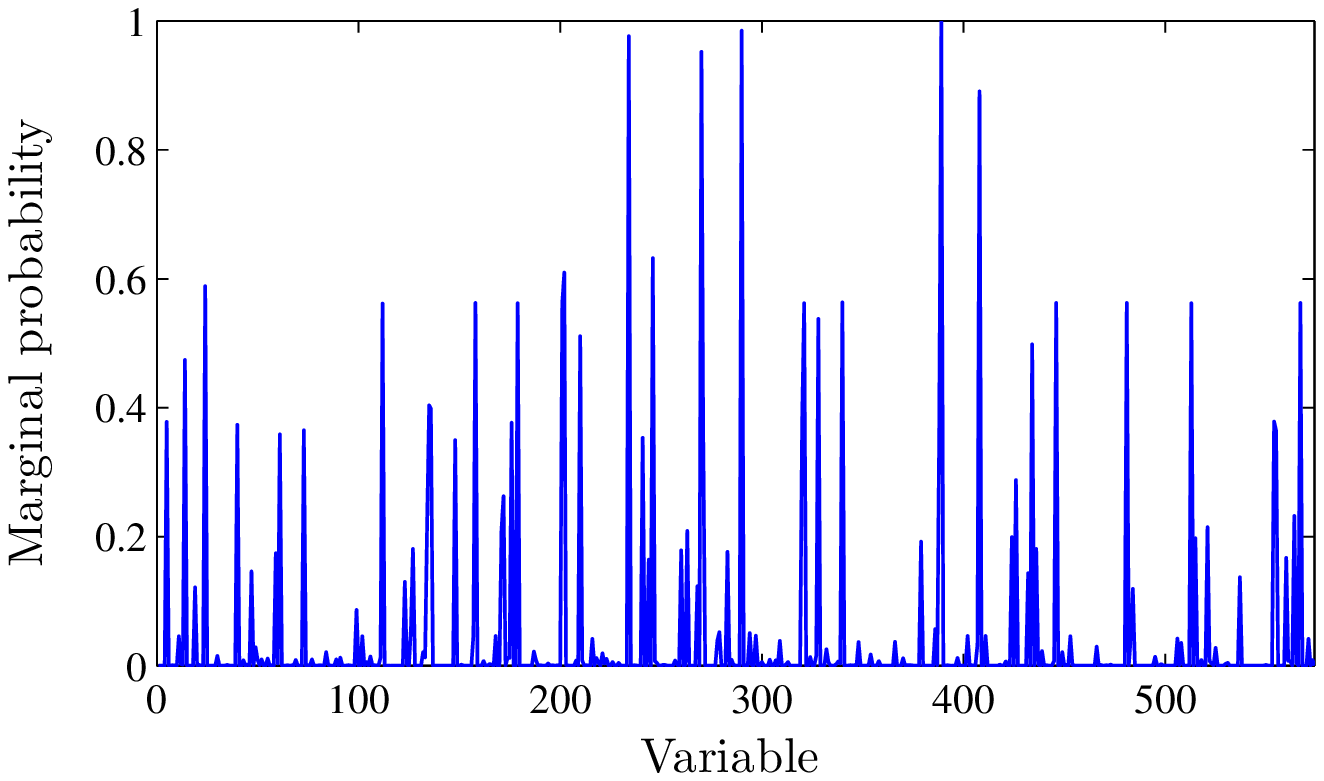}
\caption{Number of included genes (left) in each iteration and marginal posterior probabilities of inclusion (right).}
\label{fig:mclBayesian}
\end{figure}

In the final model, out of 574 genes, the Bayesian-AFT method finds only 25 genes that have the largest marginal posterior probabilities (we choose 0.38 to be the cut--off) (see also Figure \ref{fig:mclBayesian}). The left panel of Figure \ref{fig:mclBayesian} shows that MCMC chains mostly visited models with 20 to 40 genes. The right panel of Figure \ref{fig:mclBayesian} shows that not many genes have high marginal probabilities (only 25 genes with marginal probabilities greater than 0.38).

There are four genes with uniqid 24761, 27434, 27844 and 29888 that are identified by all four proposed methods and there are another five genes with uniqid 22136, 24383, 29876, 30034 and 33704 that are identified by three of the proposed methods. The overall analysis of the MCL data suggests that all four proposed methods are capable of identifying sets of genes that are potentially related to the response variable. In the analysis the AEnetCC, as in the simulations with $r_{ij}=0.5$, selects a smaller number of genes than do the other methods. However with gene expression data, a smaller number of identified genes means a more focused hypothesis for future confirmations studies, and is thus usually preferred.

\begin{figure}[ht]
\centering
\includegraphics [scale=0.450] {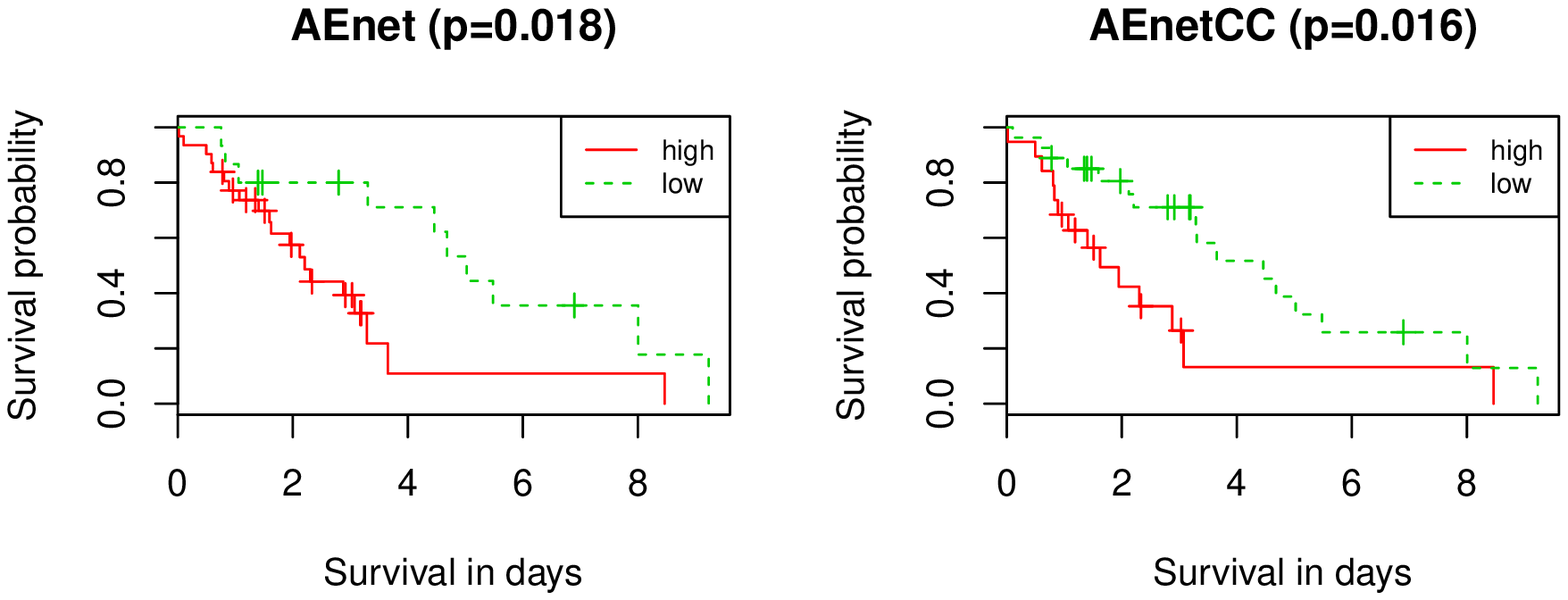}\\\includegraphics [scale=0.450] {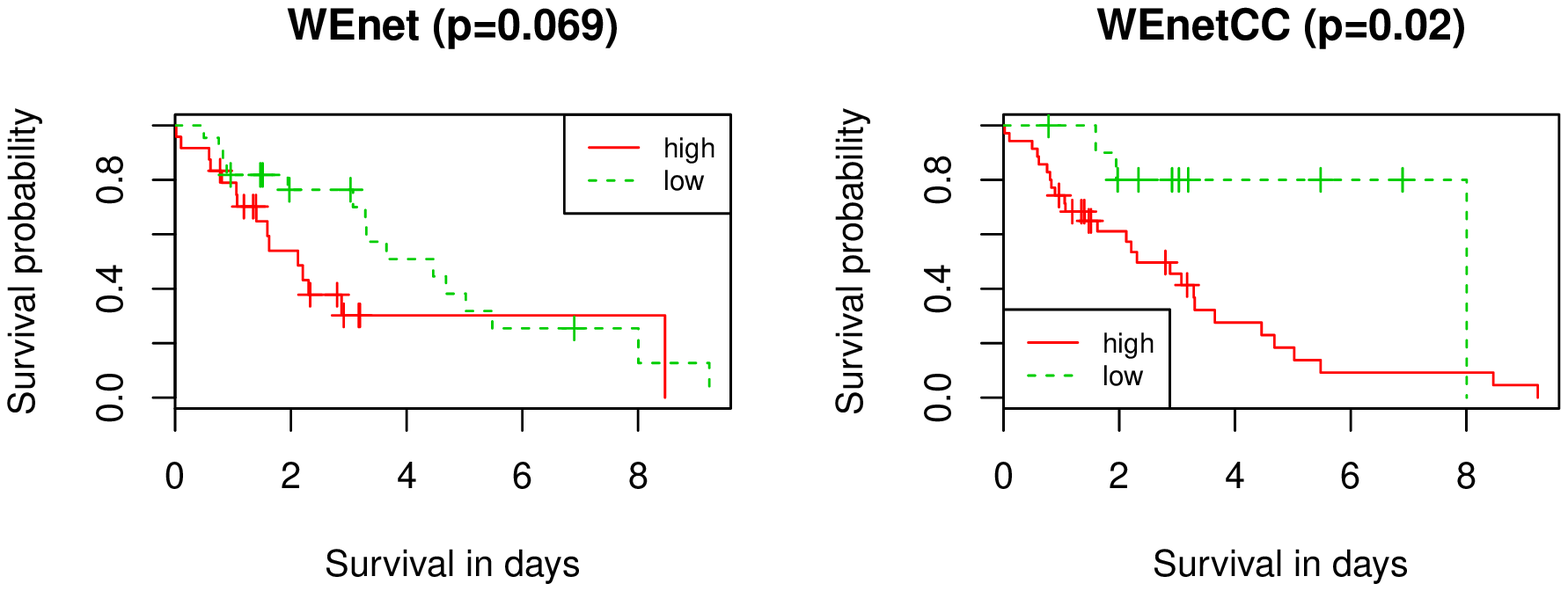}
\caption{Survival comparison between the high risk group and low risk group using different methods.}
\label{fig:mcl.predtest}
\end{figure}

We evaluate the predictive performance using the four proposed methods. We use the obtained models to predict the risk of death in the MCL test dataset. We first partition the data randomly into two equal parts called training and test datasets. We then implement the methods to the training dataset and compute the risk scores $(X^T\hat{\beta})$ based on the model estimates and the test dataset. The subjects are classified to be in the high risk group or low risk group based on whether the risk score exceeds the median survival time in the training dataset. We compare the K--M curves between the two groups and then a log--rank test is used to identify the difference between the two K--M curves (see Figure \ref{fig:mcl.predtest}). The corresponding predictive MSE for the methods AEnet, AEnetCC, WEnet, and WEnetCC are 15.7, 19.6, 29.4, and 15.0 respectively.
The log--rank test suggests that the high and low risk groups are significantly different from each other under almost all the methods. So it seems the methods can group very well the subjects' survival time into two risk sets.

%
%
\subsubsection{Mantle cell lymphoma data under adaptive preprocessing}

\begin{table}[ht]
\centering \caption{Number of genes selected by the SIS followed by the methods (diagonal elements) and number of common genes found between the methods (off diagonal elements).}
\scalebox{0.48}{\begin{tabular}{lcccccccccc}\toprule\toprule
Methods&SIS+AEnet&SIS+AEnetCC&SIS+WEnet&SIS+WEnetCC&SIS+Enet&SIS+Enet-AFT&SIS+Bys-AFT&SIS+SCAD&SIS+TCS&SIS+PC\\\hline
SIS+AEnet&\color{red}{05}&02&02&01&02&04&02&04&00&02 \\
SIS+AEnetCC&02&\color{red}{12}&05&03&07&07&02&03&00&02 \\
SIS+WEnet&02&05&\color{red}{10}&02&05&07&02&02&00&02\\
SIS+WEnetCC&01&03&02&\color{red}{08}&05&02&01&01&00&01\\
SIS+Enet&02&07&05&05&\color{red}{32}&13&02&03&01&02\\
SIS+Enet-AFT&04&07&07&02&13&\color{red}{30}&02&04&00&02\\
SIS+Bys-AFT&02&02&02&01&02&02&\color{red}{02}&02&00&02\\
SIS+SCAD&04&03&02&01&03&04&02&\color{red}{05}&00&02\\
SIS+TCS&00&00&00&00&01&00&00&00&\color{red}{01}&00\\
SIS+PC&02&02&02&01&02&02&02&02&00&\color{red}{02}\\
\bottomrule\bottomrule
\end{tabular}}
\label{tab:sismclcompsummary}
\end{table}
A challenge with the MCL dataset (with ultra-high dimensionality, $p\gg n$) is that the important genes might be highly correlated with some unimportant ones; that usually increases with dimensionality. The maximum spurious correlation between a gene and the survival time also grows with dimensionality. Here we focus on a smart preprocessing technique for MCL data, that addresses this issue and also reduces circularity bias [Kriegeskorte et at.~(2009\nocite{Krie:etal:09:circular})] by reducing false discovery rate. Fan and Lv (2008\nocite{Fan:Lv:08:Sure}) introduced the SIS idea, that reduces the ultra-high dimensionality to a relatively large scale $d_n$, where $d_n < n$. In Fan and Lv (2008\nocite{Fan:Lv:08:Sure}) asymptotic theory is proved to show that, with high probability, SIS keeps all important variables with vanishing false discovery rate. Then, the lower dimension methods such as SCAD [Fan and Li (2001)\nocite{fan:li:01:varia}] can be used to estimate the sparse model. This procedure is referred to as SIS+SCAD
For MCL data we first apply SIS to reduce the dimensionality from 8810 to $d_n=[3\,n^{2/3}]=61$ and then fit the data using all ten \emph{methods} (our four proposed methods and six competitors, including the three greedy methods). We call this procedure $SIS+methods$. The results are reported in Table \ref{tab:sismclcompsummary}.

\begin{table}[ht]
\centering \caption{Number of genes selected between the methods with and without the SIS implementation.}
\scalebox{0.6}{\begin{tabular}{lcccccccccc}\toprule\toprule
Methods&AEnet&AEnetCC&WEnet&WEnetCC&Enet&Enet-AFT&Bayesian-AFT&SIS&TCS&PC-simple\\\hline
SIS+AEnet&01&01&00&00&00&00&00&01&00&00\\
SIS+AEnetCC&01&01&00&00&01&00&00&01&00&01\\
SIS+WEnet&01&01&00&00&00&01&00&01&00&00\\
SIS+WEnetCC&01&01&00&00&00&00&00&00&00&00\\
SIS+Enet&06&06&00&01&01&02&00&03&01&01\\
SIS+Enet-AFT&04&04&00&01&01&02&00&04&00&01\\
SIS+Bys-AFT&00&00&00&00&00&00&00&00&00&00\\
SIS+SCAD&01&01&00&00&01&00&00&01&00&01\\
SIS+TCS&00&00&00&00&00&00&00&00&00&00\\
SIS+PC&00&00&00&00&00&00&00&00&00&00\\
\bottomrule\bottomrule
\end{tabular}}
\label{tab:mclcompwithwithoutsis}
\end{table}

\begin{figure}[ht]
\centering
\includegraphics [scale=0.42] {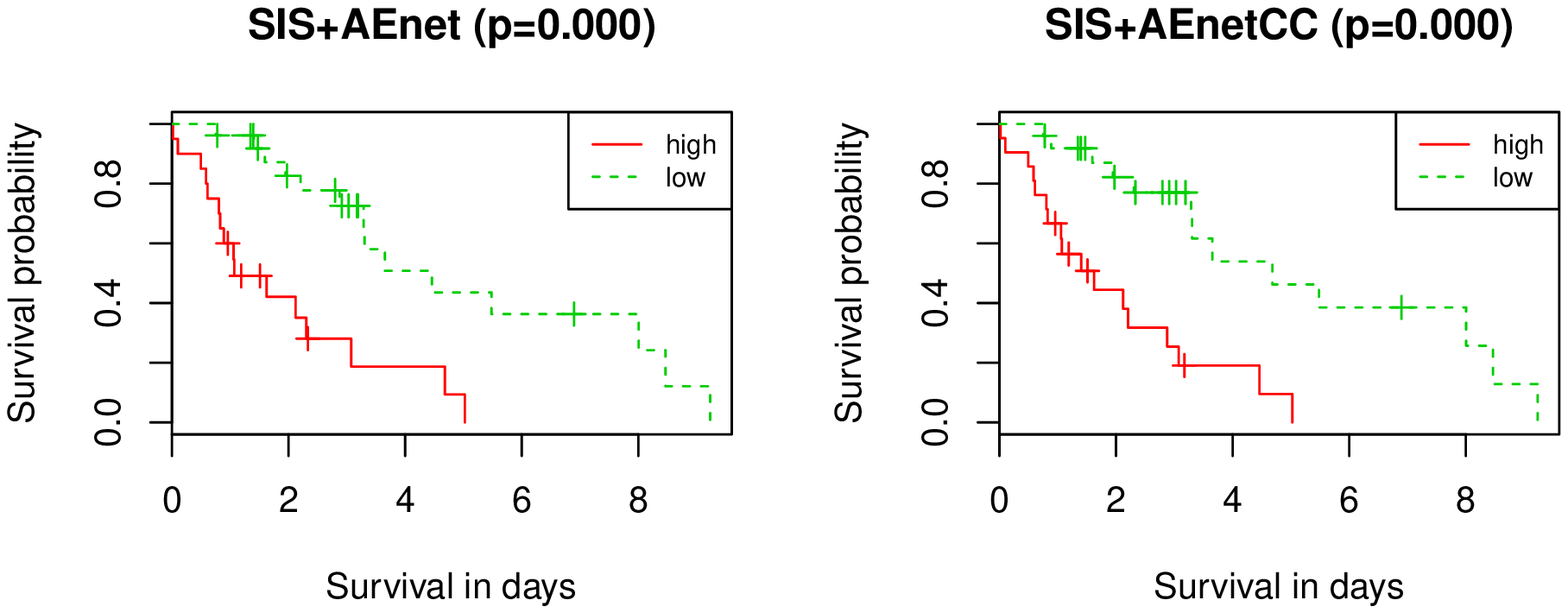}\\
\includegraphics [scale=0.42] {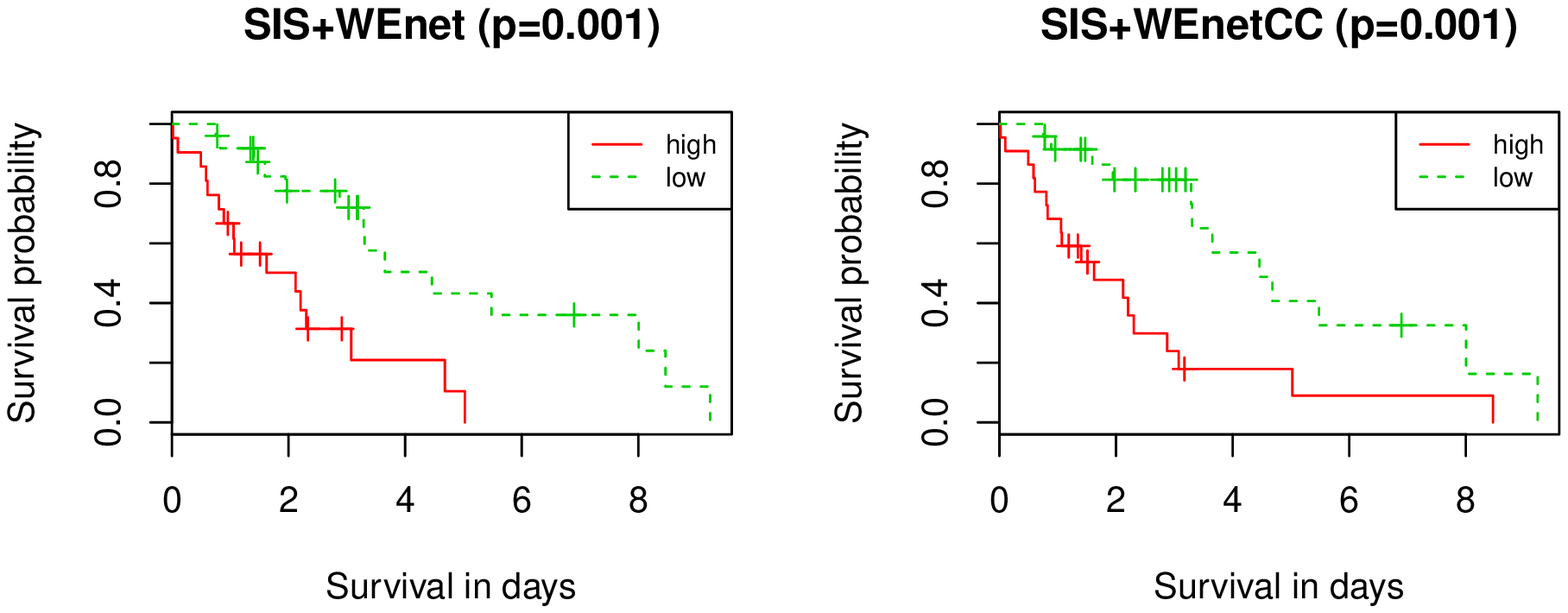}
\caption{Survival comparison between the high risk group and low risk group using different methods with SIS implementation.}
\label{fig:mcl.sis.predtest}
\end{figure}

From Table \ref{tab:sismclcompsummary} we see that the proposed methods select a considerable number of genes in common with the other methods.
The three greedy methods tend to return final models with fewer genes. TCS selects the lowest number of genes (1) while Enet selects the largest number of genes (32). Among four proposed methods, AEnet selects the lowest number of genes (5). The results also suggest that the implementation of SIS followed by all the methods (proposed and competitors) pick smaller sets of genes, most of which are not in the set of genes found by the methods without SIS. Table \ref{tab:mclcompwithwithoutsis} shows the number of common genes between the methods with and without SIS implementation. The predictive performance for the four proposed methods with SIS implementation has been evaluated (see Figure \ref{fig:mcl.sis.predtest}) similarly to what was done before for methods without SIS (Figure \ref{fig:mcl.predtest}). The predictive MSE for methods SIS+AEnet, SIS+AEnetCC, SIS+WEnet, and SIS+WEnetCC are 1.2, 1.1, 2.8, and 1.4 respectively. It is clear from the predictive performance graph Figure \ref{fig:mcl.sis.predtest} (also Figure \ref{fig:mcl.predtest} for methods without SIS) and the predictive MSE's that the predictive performance improves considerably after implementation of SIS.

\section{Discussion}
In this study we propose adaptive elastic net and weighted elastic net regularized variable selection approaches for the AFT model. We conjecture that the proposed approaches enjoy oracle properties under some regularity assumptions in analogous with the adaptive elastic net [Ghosh (2007\nocite{ghosh:07:adapEnet})] and weighted elastic net [Hong and Zhang (2010\nocite{Hon:Zhang:2010:weighted})]. They produce sparse solutions. We propose another two variable selection algorithms, where censored observations are used as extra constraints in the optimization function of the methods. The censoring constraints in the optimization equations limit the model space using the right censored data.
It is shown how all the methods apart from the AEnetCC can be optimized after transforming them into an adaptive lasso problem in some augmented space.

The analysis of both simulated and MCL gene expression data shows that the regularized SWLS approach for variable selection with its four implementations (AEnet, AEnetCC, WEnet and WEnetCC) can be used for selecting important variables. They also can be used for future prediction for survival time under AFT models. The MCL gene expression data analysis also suggests that the sure independence screening improves the performance of all the proposed methods in the AFT model. It is observed that the methods AEnetCC and WEnetCC seem only to perform well under moderately high--dimensional censored datasets such as with variables at most four or five times higher than the sample size. However, the two methods with SIS implementation have no such limits. A variable selection strategy such as ours, that allows the adaptive elastic net [Ghosh (2007\nocite{ghosh:07:adapEnet})] and weighted elastic net [Hong and Zhang (2010\nocite{Hon:Zhang:2010:weighted})] to be used for censored data, is new. The extensions of these methods, that use right censored data to limit the model space and improve the parameter estimation, are also new.
Both the adaptive and weighted elastic net together with two extensions enjoy the computational advantages of the lasso. The methods show that various regularized technique will continue to be an important tool in variable selection with survival data. Our new variable selection techniques for high--dimensional censored data are promising alternatives to the existing methods. For implementing all proposed methods we have provided a publicly available package \emph{AdapEnetClass} (Khan \& Shaw, 2013\nocite{has:ewa:Rpack:AdapEnetClass}) implemented in the R programming system.

\section*{Acknowledgements}
The first author is grateful to the centre for research in Statistical Methodology (CRiSM), Department of Statistics, University of Warwick, UK for offering research funding for his PhD study.
\bibliography{b2ndyear}
\bibliographystyle{imsart-nameyear}

\end{document}